\documentclass[reqno,12pt,a4paper]{amsart}

\usepackage{mathrsfs}
\usepackage{amssymb}
\usepackage{amsfonts}
\usepackage{latexsym}
\usepackage{amsthm}

\usepackage{graphicx}
\def\lb{\label}

\newcommand{\er}[1]{\textrm{(\ref{#1})}}

\begin{document}


\renewcommand{\theequation}{\arabic{section}.\arabic{equation}}
\theoremstyle{plain}
\newtheorem{theorem}{\bf Theorem}[section]
\newtheorem{lemma}[theorem]{\bf Lemma}
\newtheorem{corollary}[theorem]{\bf Corollary}
\newtheorem{proposition}[theorem]{\bf Proposition}
\newtheorem{definition}[theorem]{\bf Definition}
\newtheorem{remark}[theorem]{\it Remark}

\def\a{\alpha}  \def\cA{{\mathcal A}}     \def\bA{{\bf A}}  \def\mA{{\mathscr A}}
\def\b{\beta}   \def\cB{{\mathcal B}}     \def\bB{{\bf B}}  \def\mB{{\mathscr B}}
\def\g{\gamma}  \def\cC{{\mathcal C}}     \def\bC{{\bf C}}  \def\mC{{\mathscr C}}
\def\G{\Gamma}  \def\cD{{\mathcal D}}     \def\bD{{\bf D}}  \def\mD{{\mathscr D}}
\def\d{\delta}  \def\cE{{\mathcal E}}     \def\bE{{\bf E}}  \def\mE{{\mathscr E}}
\def\D{\Delta}  \def\cF{{\mathcal F}}     \def\bF{{\bf F}}  \def\mF{{\mathscr F}}
\def\c{\chi}    \def\cG{{\mathcal G}}     \def\bG{{\bf G}}  \def\mG{{\mathscr G}}
\def\z{\zeta}   \def\cH{{\mathcal H}}     \def\bH{{\bf H}}  \def\mH{{\mathscr H}}
\def\e{\eta}    \def\cI{{\mathcal I}}     \def\bI{{\bf I}}  \def\mI{{\mathscr I}}
\def\p{\psi}    \def\cJ{{\mathcal J}}     \def\bJ{{\bf J}}  \def\mJ{{\mathscr J}}
\def\vT{\Theta} \def\cK{{\mathcal K}}     \def\bK{{\bf K}}  \def\mK{{\mathscr K}}
\def\k{\kappa}  \def\cL{{\mathcal L}}     \def\bL{{\bf L}}  \def\mL{{\mathscr L}}
\def\l{\lambda} \def\cM{{\mathcal M}}     \def\bM{{\bf M}}  \def\mM{{\mathscr M}}
\def\L{\Lambda} \def\cN{{\mathcal N}}     \def\bN{{\bf N}}  \def\mN{{\mathscr N}}
\def\m{\mu}     \def\cO{{\mathcal O}}     \def\bO{{\bf O}}  \def\mO{{\mathscr O}}
\def\n{\nu}     \def\cP{{\mathcal P}}     \def\bP{{\bf P}}  \def\mP{{\mathscr P}}
\def\r{\rho}    \def\cQ{{\mathcal Q}}     \def\bQ{{\bf Q}}  \def\mQ{{\mathscr Q}}
\def\s{\sigma}  \def\cR{{\mathcal R}}     \def\bR{{\bf R}}  \def\mR{{\mathscr R}}
\def\S{\Sigma}  \def\cS{{\mathcal S}}     \def\bS{{\bf S}}  \def\mS{{\mathscr S}}
\def\t{\tau}    \def\cT{{\mathcal T}}     \def\bT{{\bf T}}  \def\mT{{\mathscr T}}
\def\f{\phi}    \def\cU{{\mathcal U}}     \def\bU{{\bf U}}  \def\mU{{\mathscr U}}
\def\F{\Phi}    \def\cV{{\mathcal V}}     \def\bV{{\bf V}}  \def\mV{{\mathscr V}}
\def\P{\Psi}    \def\cW{{\mathcal W}}     \def\bW{{\bf W}}  \def\mW{{\mathscr W}}
\def\o{\omega}  \def\cX{{\mathcal X}}     \def\bX{{\bf X}}  \def\mX{{\mathscr X}}
\def\x{\xi}     \def\cY{{\mathcal Y}}     \def\bY{{\bf Y}}  \def\mY{{\mathscr Y}}
\def\X{\Xi}     \def\cZ{{\mathcal Z}}     \def\bZ{{\bf Z}}  \def\mZ{{\mathscr Z}}
\def\O{\Omega}

\newcommand{\gA}{\mathfrak{A}}
\newcommand{\gB}{\mathfrak{B}}
\newcommand{\gC}{\mathfrak{C}}
\newcommand{\gD}{\mathfrak{D}}
\newcommand{\gE}{\mathfrak{E}}
\newcommand{\gF}{\mathfrak{F}}
\newcommand{\gG}{\mathfrak{G}}
\newcommand{\gH}{\mathfrak{H}}
\newcommand{\gI}{\mathfrak{I}}
\newcommand{\gJ}{\mathfrak{J}}
\newcommand{\gK}{\mathfrak{K}}
\newcommand{\gL}{\mathfrak{L}}
\newcommand{\gM}{\mathfrak{M}}
\newcommand{\gN}{\mathfrak{N}}
\newcommand{\gO}{\mathfrak{O}}
\newcommand{\gP}{\mathfrak{P}}
\newcommand{\gR}{\mathfrak{R}}
\newcommand{\gS}{\mathfrak{S}}
\newcommand{\gT}{\mathfrak{T}}
\newcommand{\gU}{\mathfrak{U}}
\newcommand{\gV}{\mathfrak{V}}
\newcommand{\gW}{\mathfrak{W}}
\newcommand{\gX}{\mathfrak{X}}
\newcommand{\gY}{\mathfrak{Y}}
\newcommand{\gZ}{\mathfrak{Z}}

\def\ve{\varepsilon}   \def\vt{\vartheta}    \def\vp{\varphi}    \def\vk{\varkappa}

\def\Z{{\mathbb Z}}    \def\R{{\mathbb R}}   \def\C{{\mathbb C}}
\def\T{{\mathbb T}}    \def\N{{\mathbb N}}   \def\dD{{\mathbb D}}


\def\la{\leftarrow}              \def\ra{\rightarrow}            \def\Ra{\Rightarrow}
\def\ua{\uparrow}                \def\da{\downarrow}
\def\lra{\leftrightarrow}        \def\Lra{\Leftrightarrow}


\def\lt{\biggl}                  \def\rt{\biggr}
\def\ol{\overline}               \def\wt{\widetilde}
\def\no{\noindent}


\let\ge\geqslant                 \let\le\leqslant
\def\lan{\langle}                \def\ran{\rangle}
\def\/{\over}                    \def\iy{\infty}
\def\sm{\setminus}               \def\es{\emptyset}
\def\ss{\subset}                 \def\ts{\times}
\def\pa{\partial}                \def\os{\oplus}
\def\om{\ominus}                 \def\ev{\equiv}
\def\iint{\int\!\!\!\int}        \def\iintt{\mathop{\int\!\!\int\!\!\dots\!\!\int}\limits}
\def\el2{\ell^{\,2}}             \def\1{1\!\!1}
\def\sh{\sharp}
\def\wh{\widehat}

\def\all{\mathop{\mathrm{all}}\nolimits}
\def\where{\mathop{\mathrm{where}}\nolimits}
\def\as{\mathop{\mathrm{as}}\nolimits}
\def\Area{\mathop{\mathrm{Area}}\nolimits}
\def\arg{\mathop{\mathrm{arg}}\nolimits}
\def\const{\mathop{\mathrm{const}}\nolimits}
\def\det{\mathop{\mathrm{det}}\nolimits}
\def\diag{\mathop{\mathrm{diag}}\nolimits}
\def\diam{\mathop{\mathrm{diam}}\nolimits}
\def\dim{\mathop{\mathrm{dim}}\nolimits}
\def\dist{\mathop{\mathrm{dist}}\nolimits}
\def\Im{\mathop{\mathrm{Im}}\nolimits}
\def\Iso{\mathop{\mathrm{Iso}}\nolimits}
\def\Ker{\mathop{\mathrm{Ker}}\nolimits}
\def\Lip{\mathop{\mathrm{Lip}}\nolimits}
\def\rank{\mathop{\mathrm{rank}}\limits}
\def\Ran{\mathop{\mathrm{Ran}}\nolimits}
\def\Re{\mathop{\mathrm{Re}}\nolimits}
\def\Res{\mathop{\mathrm{Res}}\nolimits}
\def\res{\mathop{\mathrm{res}}\limits}
\def\sign{\mathop{\mathrm{sign}}\nolimits}
\def\span{\mathop{\mathrm{span}}\nolimits}
\def\supp{\mathop{\mathrm{supp}}\nolimits}
\def\Tr{\mathop{\mathrm{Tr}}\nolimits}
\def\BBox{\hspace{1mm}\vrule height6pt width5.5pt depth0pt \hspace{6pt}}


\newcommand\nh[2]{\widehat{#1}\vphantom{#1}^{(#2)}}
\def\dia{\diamond}

\def\Oplus{\bigoplus\nolimits}



\def\qqq{\qquad}
\def\qq{\quad}
\let\ge\geqslant
\let\le\leqslant
\let\geq\geqslant
\let\leq\leqslant
\newcommand{\ca}{\begin{cases}}
\newcommand{\ac}{\end{cases}}
\newcommand{\ma}{\begin{pmatrix}}
\newcommand{\am}{\end{pmatrix}}
\renewcommand{\[}{\begin{equation}}
\renewcommand{\]}{\end{equation}}
\def\bu{\bullet}

\title[{ Schr\"odinger  operator with periodic plus
compactly supported potentials}]
{Schr\"odinger operator with periodic plus
compactly supported potentials on the half-line}

\date{\today}
\author[Evgeny Korotyaev]{Evgeny Korotyaev}
\address{School of Math., Cardiff University.
Senghennydd Road, CF24 4AG Cardiff, Wales, UK.
email: \ KorotyaevE@cf.ac.uk}



\begin{abstract}
\no We consider the Schr\"odinger operator $H$ with a periodic potential $p$
plus a compactly supported  potential $q$ on the half-line. We prove the following results:
1) a forbidden domain for the resonances is specified,
2) asymptotics of the resonance-counting function  is determined,
3) in each nondegenerate gap $\g_n$ for $n$ large enough there is exactly an eigenvalue  or an antibound state, 4) the asymptotics of  eigenvalues and antibound states are determined at high energy,
5) the number of eigenvalues plus antibound states is odd $\ge 1$ in each gap,
 6) between any two eigenvalues there is an odd number $\ge 1$ of antibound states,  7) for any potential $q$ and for any
  sequences $(\s_n)_{1}^\iy, \s_n\in \{0,1\}$ and $(\vk_n)_1^\iy\in \ell^2, \vk_n\ge 0$, there exists a potential $p$  such that each gap length  $|\g_n|=\vk_n, n\ge 1$ and $H$ has exactly $\s_n$ eigenvalues and $1-\s_n$ antibound state in each gap $\g_n\ne \es$ for $n$ large enough,
8) if unperturbed operator (at $q=0$) has infinitely many virtual states, then
for any sequence $(\s)_1^\iy, \s_n\in \{0,1\}$, there exists a potential $q$ such that $H$ has $\s_n$ bound states and $1-\s_n$ antibound states in each gap open $\g_n$ for $n$ large enough.

\end{abstract}

\maketitle

\section {Introduction and main results}
\setcounter{equation}{0}

 Consider the Schr\"odinger operator $H$ acting in the Hilbert space $L^2(\R_+ )$ and given by
\begin{multline}
\lb{0}
H=H_0+q,\qqq
H_0f=-f''+p(x)f,  \qq f(0)=0,\\
p\in L_{real}^1(\R/\Z),\qqq q\in \cQ_t=\{q: \ q\in L_{real}^2(\R_+), \  \sup ( \supp (q))=t\},\ t>0.
\end{multline}
The spectrum of $H_0$ consists of an absolutely continuous part $\s_{ac}(H_0)=
\cup_{n\ge 1} \gS_n$ plus  at most one eigenvalue in each gap $\g_n\ne \es, n\ge 1$ (\cite{CL}, \cite{E}, \cite{Zh3}),
 where the bands $\gS_n$ and gaps $\g_n$ are given by (see Fig. 1)
$$
\gS_n=[E^+_{n-1},E^-_n],\ \ \qq \g_{n}=(E^-_{n},E^+_n),\qqq n\ge 1,\qq
E_0^+<..\le E^+_{n-1}< E^-_n \le E^+_{n}<...
$$
Let below $E_0^+=0$. The sequence $E_0^+<E_1^-\le E_1^+\ <\dots$  is the spectrum of the equation
\[
\lb{1}
-y''+p(x)y=\l y, \ \ \ \ \l\in \C ,
\]
with the 2-periodic boundary conditions, i.e. $y(x+2)=y(x), x\in \R$.
The bands $\gS_n, \gS_{n+1}$ are separated by  a gap $\g_{n}$ and let $\g_0=(-\iy,E_0^+)$.
If a gap degenerates, that is $\g_n=\es $, then the corresponding bands $\gS_{n} $ and $\gS_{n+1}$ merge.
If $E_n^-=E_n^+$ for some $n$, then this number $E_n^{\pm}$ is the double eigenvalue of equation \er{1} with the 2-periodic boundary conditions. The lowest eigenvalue $E_0^+=0$ is always simple  and the corresponding eigenfunction is 1-periodic. The eigenfunctions, corresponding to the eigenvalue $E_{2n}^{\pm}$, are 1-periodic, and for the case $E_{2n+1}^{\pm}$ they are anti-periodic,  i.e., $y(x+1)=-y(x),\ \ x\in\R$.

We describe properties the operator $\mH y=-f''+(p+q)y$ on the real line, where $p$ is periodic and and $q$ is compactly supported. The spectrum of $\mH$ consists of an absolutely continuous part $\s_{ac}(\mH)=\s_{ac}(H_0)$ plus a finite number of simple eigenvalues in each gap $\g_n\ne \es, n\geq 0$, see  \cite{Rb}, \cite{F1} and at most two eigenvalue \cite{Rb} in every open gap $\g_n$ for $n$ large enough.
If $q_0=\int_\R q(x)dx\ne 0$, then $\mH$ has  precisely one eigenvalue (see \cite{Zh1}, \cite{F2}, \cite{GS}) and one antibound state \cite{K4} in each gap $\g_n\ne \es$  for $n$ large enough.
If  $q_0=0$, then roughly speaking  there are two eigenvalues and zero antibound state or zero eigenvalues and two antibound states \cite{K4}
in each gap $\g_n\ne \es$  for $n$ large enough.

The spectrum of $H$ acting in $L^2(\R_+)$ consists of an absolutely continuous part $\s_{ac}(H)=\s_{ac}(H_0)$ plus a finite number of simple eigenvalues in each gap $\g_n\ne \es, n\geq 0$. Note that the last fact follows from the same result for $\mH$ and the splitting principle.

\begin{figure}
\tiny
\unitlength=1.00mm
\special{em:linewidth 0.4pt}
\linethickness{0.4pt}
\begin{picture}(108.67,33.67)
\put(41.00,17.33){\line(1,0){67.67}}
\put(44.33,9.00){\line(0,1){24.67}}
\put(108.33,14.00){\makebox(0,0)[cc]{$\Re\l$}}
\put(41.66,33.67){\makebox(0,0)[cc]{$\Im\l$}}
\put(42.00,14.33){\makebox(0,0)[cc]{$0$}}
\put(44.33,17.33){\linethickness{4.0pt}\line(1,0){11.33}}
\put(66.66,17.33){\linethickness{4.0pt}\line(1,0){11.67}}
\put(82.00,17.33){\linethickness{4.0pt}\line(1,0){12.00}}
\put(95.66,17.33){\linethickness{4.0pt}\line(1,0){11.00}}
\put(46.66,20.00){\makebox(0,0)[cc]{$E_0^+$}}
\put(56.66,20.33){\makebox(0,0)[cc]{$E_1^-$}}
\put(68.66,20.33){\makebox(0,0)[cc]{$E_1^+$}}
\put(78.33,20.33){\makebox(0,0)[cc]{$E_2^-$}}
\put(84.33,20.33){\makebox(0,0)[cc]{$E_2^+$}}
\put(93.00,20.33){\makebox(0,0)[cc]{$E_3^-$}}
\put(98.66,20.33){\makebox(0,0)[cc]{$E_3^+$}}
\put(106.33,20.33){\makebox(0,0)[cc]{$E_4^-$}}
\end{picture}
\caption{The cut domain $\C\sm \cup \gS_n$ and the cuts (bands) $\gS_n=[E^+_{n-1},E^-_n], n\ge 1$}
\lb{sS}
\end{figure}

Let $\vp(x,z), \vt(x,z)$ be the solutions of the equation $-y''+py=z^2y$
satisfying $\vp'(0,z)=\vt(0,z)=1$ and $\vp(0,z)=\vt'(0,z)=0$, where $y'=\pa_x y$.
The Lyapunov function is defined by $\D(z)={1\/2}(\vp'(1,z)+\vt(1,z))$.
The function $\D^2(\sqrt \l)$ is entire, where $\sqrt \l$ is defined by $\sqrt 1=1$. Introduce the function $(1-\D^2(\sqrt \l))^{1\/2}, \l\in \ol\C_+$ and we fix the branch  by the condition
$(1-\D^2(\sqrt {\l+i0}))^{1\/2}>0$ for $\l\in \gS_1=[E^+_{0},E^-_1]$.
Introduce the two-sheeted Riemann surface $\L$ of $(1-\D^2(\sqrt \l))^{1\/2}$ obtained by joining the upper and lower rims of two copies of the cut plane $\C\sm\s_{ac}(H_0)$ in the usual (crosswise) way.
The n-th gap on the first physical sheet $\L_1$ we will denote by $\g_n^{(1)}$ and the same gap but on the second
 nonphysical sheet $\L_2$ we will denote by $\g_n^{(2)}$ and let $\g_n^c$ be the union of $\ol\g_n^{(1)}$ and $\ol\g_n^{(2)}$:
\[
\lb{sL}
 \g_n^c=\ol\g_n^{(1)}\cup \ol\g_n^{(2)}.
\]

It is well known that the function $f(\l)=((H-\l)^{-1}h,h)$ has meromorphic extension from the physical sheet $\L_1$ into the Riemann surface $\L$ for each $h\in C_0^\iy(\R_+), h\ne 0$. Moreover, if $f$ has a pole at some $\l_0\in \L_1$ and some $h$, then $\l_0$ is an eigenvalue of $H$ and $\l_0\in\cup \g_n^{(1)}$.

\no {\bf Definition $\L$. } {\it Let $f(\l)=((H-\l)^{-1}h,h), \l\in \L$ for some    $h\in C_0^\iy(\R_+), h\ne 0$.

\no 1) If  $f(\l)$ has a pole at some $\l_0\in \L_2$,   $\l_0\ne E_n^\pm, n\ge 0$, we say that $\l_0$ is a resonance.

\no 2)  A point  $\l_0=E_n^\pm , n\ge 0$ is a virtual state, if  the function $f(\l_0+z^2)$ has a pole at $0$.

\no 3) A point $\l\in\L$ is a state, if it is a bound state or a resonance or a virtual state. Let $\gS_{st}(H)$ be the set of all states. The multiplicity of the state is the multiplicity of the corresponding pole.
If $\l_0\in \g_n^{(2)}, n\ge 0$, then we call $\l_0$ an antibound state.
}

As a good example we consider the states of the operator $H_0$ for  the case $p\ne \const , q=0$, see \cite{Zh3}, \cite{HKS}. Let $f_0(\l)=((H_0-\l)^{-1}h,h)$ for some $h\in C_0^\iy(\R_+)$. It is well known that the function  $f_0$ is meromorphic on the physical sheet $\L_1$ and has a meromorphic extension  into $\L$. For each $\g_n^c\ne \es, n\ge 1$ there is exactly one state  $\l_n^0\in \g_n^c$ of $H_0$ and its projection on the complex plane coincides with the Dirichlet eigenvalues $\m_n^2$. Moreover, there is one case from three ones:

1) $\l_n^0\in \g_n^{(1)}$ is an eigenvalues,

2) $\l_n^0\in \g_n^{(2)}$ is an antibound state,

3) $\l_n^0=E_n^+$ or $\l_n^0=E_n^-$ is a virtual states.

\no There are no other states of $H_0$. Thus $H_0$ has only eigenvalues, virtual states and antibound states. If there are exactly $N\ge 1$ nondegenerate gaps in the spectrum of $\s_{ac}(H_0)$, then operator
$H_0$ has exactly $N$ states. The gaps $\g_n=\es$ do not give contribution to the states. In particular, if all $\g_n=\es, n\ge 1$, then $p=0$, see \cite{MO} or \cite{K4} and $H_0$ has not states. The states $\l_n^0$ are described in Lemma \ref{Tm}.

We need the results about the inverse spectral theory for the unperturbed operator $H_0$:
define the mapping $p\to \x=(\x_n)_1^\iy$, where the components $\x_n=(\x_{1n},\x_{2n})\in \R^2$ are given by
$$
 \x_{1n}={E_n^-+E_n^+\/2}-\m_n^2,\qqq
\x_{2n}=\rt|{|\g_n|^2\/4}-\x_{1n}^2\rt|^{1\/2}a_n,\
\qq
a_n=\ca +1  & if \ \ \l_n^0\ {\rm is \ an \ eigenvalues} \\
        -1 & if \ \ \l_n^0 \ \ \ {\rm is \ a \ resonance}      \\
         0 & if  \ \ \l_n^0 \ \ {\rm is \ a \ virtual \ state}  \ac.
$$
Recall the results from \cite{K5}: {\it The mapping $\x: \cH\to \ell^2\os \ell^2$ is a real analytic isomorphism between real Hilbert spaces $\cH=\{p\in L^2(0,1): \int_0^1p(x)dx=0\}$ and $\ell^2\os \ell^2$ and the estimates hold true
\[
\lb{esg}
\|p\|\le 4\|\x\| (1+ \|\x\|^{1\/3}),\qqq
\|\x\|\le \|p\|(1+\|p\|)^{1\/3},
\]
where $\|p\|^2=\int_0^1p^2(x)dx$ and $\|\x\|^2={1\/4}\sum |\g_n|^2$.} Estimates \er{esg} were proved in \cite{K7}.
Moreover, for any sequence $\vk=(\vk_n)_1^\iy\in \ell^2, \vk_n\ge 0$ there are unique 2 -periodic eigenvalues
$E_n^\pm, n\ge 0$ for some $p\in \cH$ such that each $\vk_n=E_n^+-E_n^-, n\ge 1$. Thus if we know gap lengths $(|\g_n|)_1^\iy$, then we can recover the Riemann surface $\L$ uniquely plus the points $E_n^-=E_n^+$, if $\vk_n=0$.
Furthermore, for any sequence $\wt\l_n^0\in \g_n^c, n\ge 1$, there is an unique potential $p\in \cH$, such that each state $\l_n^0$ (corresponding to $p$) coincides with $\wt\l_n^0, n\ge 1$. Remark that results of \cite{K5} were generalized in \cite{K6} for periodic distributions $p=w'$, where $w\in\cH$.

Define the function
$$
D(\l)=\det (I+q(H_0-\l)^{-1}), \qqq \l \in \C_+.
$$
It is well known that the function $D(\l)$ is analytic in $\l \in \C_+$ and has a meromorphic extension into $\L$.
Each zero of $D(\l)$ in $\L_1$ is an eigenvalue of $H$ and belongs to the union of physical gaps $\cup_{n\ge 0} \g_n^{(1)}$. Until now only some particular results are known about the zeros on the nonphysical sheet $\L_2$. Remark that the set of zeros of $D$ on $\L_2$ is symmetric with respect to the real line, since $D$ is real on $\g_0^{(2)}$.

Let $\F (x,z)$ be the fundamental solution of the equation
\[
\lb{bcF}
-\F''+(p+q)\F=z^2 \F, x\ge 0,\qqq \F (0,z)=0, \ \ \ \F'(0,z)=1, \qq z\in\C.
\]

\begin{theorem}
\lb{T1}

\no i)  Let $\ve >0$.  The function $D$ satisfies
\[
\lb{T1-1}
D(\l)=1+{\wh q(z)-\wh q(0)\/2iz}+{O(e^{t (|\Im z|-\Im z)})\/z^{2}}\
\ \ as \ \ |z|\to\iy, z=\sqrt \l, \ \sqrt 1=1,
\]
 where $|\l-E_n^\pm|\ge n\ve $ for all $n\ge 1$ and $ \wh q(z)=\int_0^tq(x)e^{2izx}dx$ and
\[
\lb{st1}
\gS_{st}(H)\sm \gS_{st}(H_0)=\{ \l\in  \L\sm\gS_{st}(H_0):\ D(\l)=0\}
\ss \L_2\cup \bigcup_{n\ge 0}\g_n^{(1)},
\]
\[
\lb{st2}
\gS_{st}(H)\cap \gS_{st}(H_0)=\gS_{st}(H_0)\cap \{ z\in\s_{st}(H_0):\ \F(n_t,z)=0 \},\qq  n_t=\inf_{n\in \N, n\ge t} n.
\]

\no ii) If $\l_n^0\in  \gS_{st}(H)\cap \gS_{st}(H_0)$, then
$\l_n^0\in \ol\g_n^{(j)}\ne \es$ for some $j=1,2,n\ge1$ and
\[
\lb{ajf}
D(\l)\to D(\l_n^0)\ne 0\qq  as \qq \l\to \l_n^0.
\]
\no iii) (The logarithmic low.) Each resonance $\l\in \L_2$ of $H$  satisfies
\[
\lb{T1-2}
|\sqrt \l\sin \sqrt \l|\le C_Fe^{(2t+1)|\Im \sqrt \l|},\qqq
\ \ C_F=3(\|p\|_1+\|p+q\|_t)e^{2\|p+q\|_t+\|p\|_1},
\]
and there are no resonaces in the domain $\mD_{forb}=\{\l\in \L_2\sm \cup \ol\g_n^{(2)}: 4C_Fe^{2|\Im \sqrt \l|}<|\l|^{1\/2}\}$.
\end{theorem}
\no {\bf Remark.} 1)    Let $\l_n^0\in \g_n^{(1)}$ be a eigenvalue of $H_0$ for some $n\ge 1$. If $\F(n_t,\m_n)=0$, then $\m_n^2$ is a Dirichlet eigenvalue
of the problem $-y''+(p+q)y=\m_n^2 y, y(0)=y(n_t)=0$. Then by \er{st2},    $\l_n^0$ is a bound state of $H$ and \er{ajf} yields $D(\l_n^0)\ne 0$. Thus $\l_n^0$ {\bf is a pole of a resolvent, but $\l_n^0$ is neither  a zero of $D$ nor  a pole } of the S-matrix for $H,H_0$ given by
\[
\lb{sm}
\cS_M(z)={\ol {D(\l)\/ D(\l)}},\ \ \ \l \in \s_{ac}(H_0).
\]

\no 2) If $D(\l)=0$ for some $\l=E_n^\pm\ne \m_n^2, n\ge 0$, then by \er{st1}, $\l$ is a virtual state.

\no 3) If $\m_n^2=E_n^\pm$ for some $n\ge 1$, then by \er{st2}, $\m_n^2$ is a virtual state iff $\F(n_t,\m_n)=0$.

\no 4) If $p=0$, then it is well known that each zero of $D(\cdot )$ is a state, see e.g., \cite{K1}, \cite{S}.
Moreover, each  resonance lies below a logarithmic curve  (depending only in $q$ see e.g. \cite{K1}, \cite{Z}).
 The forbidden domain $\mD_{forb}\cap \C_-$  is similar to the case $p=0$, see \cite{K1}.

Let $\#(H,r, A)$ be the total number of state of $H$ in the set
$A\subseteq \L$ having modulus $\le r$, each state being counted according to its multiplicity.

Define the Fourier coefficients $p_{sn}, \wh q_{cn}$ and the Fourier transform $\wh q$ by
\[
q_0=\!\int_0^1\!q(x)dx,\ \    p_{sn}=\!\int_0^1\!p(x)\sin 2\pi nxdx,\ \
\wh q(z)=\!\int_0^t\!q(x)e^{2izx}dx, \qq \wh q_{cn}=\Re \wh q(\pi n).
\]

\begin{theorem}
\lb{T2}
\no  i) $H$ has an odd number $\ge 1$ of states on each set
$\g_n^c\ne \es,n\ge 1$, where $\g_n^c$ is a union of the physical $\ol \g_n^{(1)}\ss \L_1$ and non-physical gap $\ol \g_n^{(2)}\ss\L_2$  and $H$ has exactly one simple state $\l_n\in \g_n^c$ for all $n\ge 1+4C_Fe^{t{\pi\/2}}$ with asymptotics
\[
\lb{T2-1}
\sqrt{\l_n}=\m_n-{(q_{0}-\wh q_{cn})p_{sn}+O({1\/n})\/2(\pi n)^2} \qqq \as \qq n\to \iy.
\]
Moreover,  the following asymptotics hold true as $r\to \iy$:
\[
\lb{T2-2}
\#(H,r, \L_2\sm \cup \g_n^{(2)})=r{2t+o(1)\/\pi},
\]
\[
\lb{T2-3}
\#(H,r, \R)=\#(H_0,r, \R)+2N_q \qqq for\ some\ integer \ N_q\ge 0, \qq r\notin \cup \ol\g_n.
\]
\no ii) Let $\l$ be an eigenvalue of $H$ and let $\l^{(2)}\in \L_2$ be the same number but on the second sheet $\L_2$. Then $\l^{(2)}$ is not an anti-bound state.

\no iii) Let $\l_1<\l_2$ and let $\l_1,\l_2\in \g_n^{(1)}$ be some eigenvalues of $H$  for some $n\ge 0$ and assume that there are no other eigenvalues on the interval $\O=(\l_1,\l_2)\ss \g_n^{(1)}$.
 Let $\O^{(2)}\ss \g_n^{(2)}\ss \L_2$ be the same interval but on the second sheet.
Then there exists an odd number $\ge 1$ of antibound states on  $\O^{(2)}$.

\end{theorem}

\no {\bf Remark.} 1) Results (iii) at $p=0$ were obtained independently in \cite{K1}, \cite{S}.

\no 2) First term in the asymptotics \er{T2-2} does not depend on the periodic potential $p$.  Recall that asymptotics \er{T2-2} for the case $p=0$ was obtained by Zworski \cite{Z}.

\no 2) The main difference between the distribution of the resonances for the case $p\ne \const$ and $p=\const$ is the bound states and antibound states in high  energy gaps, see \er{T2-1}.

\no 3)  In the proof of \er{T2-2} we use the Paley Wiener type Theorem from \cite{Fr}, the Levinson Theorem (see Sect. 4) and analysis of the function $D$ near $\l_n^0$.

\no 4) For even potentials $p\in L_{even}^2(0,1)=\{p\in L^2(0,1),
p(x)=p(1-x), x\in (0,1)\}$ all coefficients $p_{sn}=0$ and
asymptotics \er{T2-1} are not sharp. This case is described in Theorem \ref{T4}.

\no 5)  Assume that a potential $u\in L^2(\R_+)$ is compactly supported, $\supp u\ss (0,t)$  and satisfies $|\wh u_{n}|=o(n^{-\a})$ as $n\to \iy$. Then in the case (ii) the operator $H+u$ has the same number of bound states in each gap $\g_n\ne \es$ for $n$ large enough.

We consider a stability of real states $\l_n$.
Recall that $\l_n^0\in \g_n^c$ is a state of $H_0$.

\begin{theorem}
\lb{T3}
Let $b_n=q_0-\wh q_{cn}$. Assume that $|p_{sn}|>n^{-\a}$ and $|b_n|>n^{-(1-\a)}$ for some $\a\in (0,1)$
and for all $n\in \N_0$, where  $ \N_0\ss \N$ is some infinite subset such that each $|\g_n|>0, n\in \N_0$.  Let $b_n>0$ (or $b_n<0$). Then the real state $\l_n\in\g_n^c$ for $n\in \N_0$ large  enough satisfies:

\no If $\l_n^0$ is an eigenvalue of $H_0$, then $\l_n$ is a an eigenvalue of $H$ and $\l_n^0<\l_n$ (or $\l_n^0>\l_n$).

\no If $\l_n^0$ is an antibound state of $H_0$, then $\l_n$ is an antibound state of $H$ and $\l_n<\l_n^0$ (or $\l_n^0<\l_n$).

\end{theorem}

{\bf Remark.} 1) Let $q> 0$. It is well known that if the coupling constant $\t>0$ is increasing,
then eigenvalues of $H_0+\t q$  are increasing too. Roughly speaking, in our case the antibound states in the gap move in opposite direction.

\no 2)  We explain roughly transformations: resonances $\to $ antibound states
$\to $ bound states. Consider the operator
$H_\t=H_0+\t q$, where $\t\in \R$ is  the coupling constant. If $\t=0$, then $H_0$ has only states $\l_n^0, n\ge 1$ (eigenvalues, antibound states and virtual states). Consider the first gap $\g_1^c\ne \es$.
 If $\t$ is increasing, then the state $\l_1^0$ moves and there are no other
 states on $\g_1^c$. If $\t$ is increasing again, then
 $\l_1^0$ removes on the physical gap $\g_1^{(1)}$ and becomes an eigenvalue;
 there are no new eigenvalues but some two complex resonances ($\l\in \C_+\ss\L_2$ and $\ol \l\in \C_-\ss\L_2$) reach non physical gap $\g_1^{(2)}$ and transform into  new antibound states. If $\t$ is increasing again, then
some new antibound states will be virtual states, and then later they will be bound states. Thus if $\t$ runs through $\R_+$, then there is a following transformation: \ resonances $\to $ antibound states $\to $ virtual states  $\to $ bound states $\to $ virtual states...

Recall that $p\in L_{even}^2(0,1)$  iff $\m_n^2=E_n^-$  or $\m_n^2=E_n^+$  for all $n\ge 1$, see \cite{GT}, \cite{KK1}.

\begin{theorem}
\lb{T4}
\no i)  Let  unperturbed states $\l_n^0\in \{E_n^-, E_n^+\}$ for all $n\in \N_0$ large enough, where $ \N_0\ss \N$ is some infinite subset such that each $|\g_n|>0, n\in \N_0$.
Then
\[
\lb{T4-1}
\sqrt{\l_n}=\m_n+s_n|\g_n|{(q_0-\wh q_{cn}+O({1\/n}))^2\/(2\pi n)^2}, \qqq
\qqq s_n=\ca +  & if \ \ \m_n^2=E_n^- \\
        - & if \ \ \m_n^2=E_n^+  \ac,\qq n\in \N_0
\]
as   $n\to \iy$. Moreover, if  $\a\in ({1\/2},1)$, then for $n\in\N_0$ large enough the following holds true:

if $\l_n^0=E_n^-,\  q_0-\wh q_{cn}>n^{-\a}$ or $\l_n^0=E_n^+, \ q_0-\wh q_{cn}<-n^{-\a}$, then $\l_n$ is an eigenvalue,

if $\l_n^0=E_n^-, \ q_0-\wh q_{cn}<-n^{-\a}$ or  $\l_n^0=E_n^+, \
q_0-\wh q_{cn}>n^{-\a}$, then $\l_n$ is an antibound state.

\no ii) Let $q\in \cQ_t$ satisfy $|q_{0}-\wh q_{cn}|>n^{-\a}$ for $n$ large enough and for some $\a\in ({1\/2},1)$. Then for any sequences $(\s_n)_{1}^\iy, \s_n\in \{0,1\}$ and $(\vk_n)_1^\iy\in\ell^2, \vk_n\ge 0$ there exists a potential $p\in L^2(0,1)$  such that each gap length  $|\g_n|=\vk_n, n\ge 1$ and $H$ has exactly $\s_n$ eigenvalues and $1-\s_n$ antibound states in each gap $\g_n\ne \es$ for $n$ large enough.

\no iii)  Let $p\in L^1(0,1)$ and let unperturbed states $\l_n^0\in \{E_n^-, E_n^+\}$ for all $n\in \N_0$ large enough, where $ \N_0\ss \N$ is some infinite subset such that each $|\g_n|>0, n\in \N_0$. Then for any sequence $(\s_n)_{1}^\iy, \s_n\in \{0,1\}$ there exists a potential  $q\in \cQ_t$  such that $H$ has exactly $\s_n$ eigenvalues and $1-\s_n$ antibound states in each gap $\g_n\ne \es$ for $n\in \N_0$ large enough.

\end{theorem}

{\bf Remark.} Roughly speaking \er{T4-1}
is asymptotics for even potentials $p\in L^1(0,1)$.

Let $\#_{bs}(H,\O)$ (or $\#_{abs}(H,\O)$)  be the total number of bound states
 (or anti bound states) of $H$ on the segment $\O\ss \ol\g_n^{(1)}\ss\L_1$ (or $\O\ss \ol\g_n^{(2)}\ss\L_2$) for some $n\ge 0$ (each antibound state being counted according to its multiplicity).

Introduce the integrated density of states $\r(\l),\l\in \R$ (a continuous function on $\R$) by
\[
\lb{ids}
\r(\l)|_{\g_n}=n, \qqq \r(\gS_{n+1})=[n,n+1] \qqq
\cos \pi \r(\l)=\D(\sqrt \l), \ \l\in \gS_{n+1}, \  \ n\ge 0.
\]
The real function $\r$ is strongly increasing on each spectral band $\gS_n$.
It is well known  that $\r(\l)={1\/\pi}\Re k(\sqrt {\l+i0}), \l\in \R$,
where $k$ is the quasimomentum defined in Section 2.

\begin{corollary}
\lb{T5}
Let $H_\t=H_0+q_\t$ where $q_\t=q({x\/\t}), \t\ge 1$.
Let $\O=[E_1,E_2]\ss \ol\g_n^{(1)}\ne \es$ be some interval on the physical sheet $\L_1$ for some $n\ge 0$ and let $\O^{(2)}\ss \ol\g_n^{(2)}$ be the same interval, but on the non- physical sheet $\L_2$. Then
\begin{multline}
\lb{T5-1}
\#_{abs}(H_\t,\O^{(2)})\ge 1+\#_{bs}(H_\t,\O)=
\\
\t \int_0^\iy \rt(\r(E_2-q(x))-\r(E_1-q(x))\rt)dx+o(\t)\qq
\as \qqq \t\to \iy.
\end{multline}
\end{corollary}

\no {\bf Remark.} 1) In the proof of \er{T5-1} we use the Sobolev's results  \cite{So} about asymptotics  of $\#_{bs}(H_\t,\O)$ with needed modifications of Schmidt \cite{Sc}. Sobolev considered the case $H_\t=H_0+\t V$, where $V(x)={c+o(1)\/x^\a}$ as $x\to\iy$, for some $c\ne 0,\a>0$. We can not apply this results to our case, since this potential $V$ is not compactly supported. We use the modification of Schmidt for the perturbation of the periodic Dirac operator, where
the decreasing potential can be compactly supported.

A lot of papers are devoted to the resonances for the Schr\"odinger
operator with $p=0$, see \cite{Fr}, \cite{H}, \cite{K1}, \cite{K2}, \cite{S}, \cite{Z}  and references therein. Although resonances have been studied in many settings, but there are relatively few cases where the asymptotics of the resonance counting function are known, mainly one dimensional case  \cite{Fr}, \cite{K1}, \cite{K2}, \cite{S}, and \cite{Z}. We recall that
Zworski [Z] obtained the first results about the distribution of
resonances for the Schr\"odinger operator with compactly supported
potentials on the real line. The author obtained the characterization (plus uniqueness and recovering) of $S$-matrix for the Schr\"odinger operator with a compactly supported potential on the real line \cite{K2} and the half-line \cite{K1}, see also \cite{Z1}, \cite{BKW} about uniqueness.

The author \cite{K3}  obtained the stability results for the Schr\"odinger operator on the half line:

(i) If $\vk=(\vk)_1^\iy$ is a sequence of poles (eigenvalues
and resonances) of the S-matrix for some real compactly
supported potential $q$ and $\wt\vk-\vk\in\ell_\ve^2$ for some
$\ve>1$, then $\wt\vk$ is the sequence of poles of the S-matrix
for some unique real compactly supported potential $\wt q$.

(ii) The measure associated with the poles of the S-matrix is the Carleson measure, the sum $\sum (1+|\vk_n|)^{-\a}, \a>1$ is estimated in terms of the $L^1$-norm of the potential $q$.

Brown and Weikard \cite{BW} considered  the Schr\"odinger operator $-y''+(p_A+q)y$ on the half-line,
where $p_A$ is an algebro-geometric potentials
and $q$ is a compactly supported potential. They proved that the zeros of the Jost function determine $q$ uniquely.

Christiansen \cite{Ch} considered resonances associated to the Schr\"odinger
operator $-y''+(p_{S}+q)y$ on the real line, where $p_S$ is a step potential. She determined asymptotics of the resonance-counting function. Moreover, she obtained  that the resonances determine $q$ uniquely.

Describer recent author's results \cite{K4} about the operator $\mH
=\mH_0+q$, $\mH_0=-{d^2\/dx^2}+p$ on the real line, where $p$ is periodic
and and $q$ is compactly supported:
1) asymptotics of the resonance-counting function is determined,
2) a forbidden domain for the resonances is specified,
3) the asymptotics of eigenvalues and antibound states are determined,
4) for any sequence $(\s)_1^\iy, \s_n\in \{0,2\}$, there exists a compactly supported potential $q$ such that $\mH$ has $\s_n$ bound states and $2-\s_n$ antibound states in each gap $\g_n\ne \es$ for $n$ large enough,
5) for any $q$ (with $q_0=0$) and for any sequences $(\s_n)_{1}^\iy, \s_n\in \{0,2\}$ and   $(\vk_n)_1^\iy\in \ell^2, \vk_n\ge 0$
there exists a potential $p\in L^2(0,1)$  such that each gap length  $|\g_n|=\vk_n, n\ge 1$ and $\mH$ has exactly $\s_n$ eigenvalues and $2-\s_n$ antibound states in each gap $\g_n\ne \es$ for $n$ large enough.

We compare the results for $\mH$ on $\R$ and $H$ on $\R_+$:
1) their properties are close for even potentials $p\in L_{even}^2(0,1)$,
since in this case unperturbed operators $H_0$ have only virtual states,
2) if $p$ is not even, then the unperturbed operator $H_0$ (in general)
has eigenvalues, virtual states and antibound states, but the operator $\mH_0$
has exactly two virtual states in each open gap. This leads to the different
properties of $H, \mH$ and roughly speaking the case of $H$ is more complicated,
since the unperturbed operator $H_0$ is more comlicated.

The plan of the paper is as follows. In Section 2  we  define the Riemann surface associated the momentum variable
$z=\sqrt \l, \l\in \L$, and describe the preliminary results about fundamental solutions. In Sections 3 we study states of $H$. In Sections 4 we prove the main Theorem \ref{T1}-\ref{T4}.
In the proof of theorems we use properties of the quasimomentum,
a priori estimates from \cite{KK}, \cite{MO},   and
results from the inverse theory for the Hill operator from \cite{K5}.
In the proof the analysis of the function $F(z)=\vp(1,z)D(z)\ol D(z), z^2\in \s_{ac}(H) $ is important, since we obtain the relationship between
zeros of $F$ (which is entire) and the states.
Thus we reduce the spectral problems of $H$ to the problem of entire
function theory.


\section {Preliminaries }
\setcounter{equation}{0}

We will work with the momentum $z=\sqrt \l$, where $\l$ is an energy and recall
that $\sqrt 1=1$.  Introduce the cut domain (see Fig.2)
\[
\lb{2}
\cZ=\C\sm \cup \ol g_n,\qqq where \qq
g_n=(e_n^-,e_n^+)=-g_{-n},\qq e_n^\pm=\sqrt{E_n^\pm}>0,\qq
n\ge 1.
\]
Note that $\D(e_{n}^{\pm})=(-1)^n$. If $\l\in \g_n,  n\ge 1$, then $z\in g_{\pm n}$ and if $\l\in \g_0=(-\iy,0)$, then  $z\in g_0^\pm=i\R_\pm$. Below we will use the momentum variable $z=\sqrt \l$ and the corresponding Riemann surface $\cM$, which is more convenient for us, than the Riemann surface $\L$. Slitting the n-th momentum gap  $g_n$ (suppose it is nontrivial) we obtain a cut $g_n^c$ with an upper   $g_n^+$ and lower rim  $g_n^-$. Below we will identify this cut $g_n^c$ and the union of of the upper rim  (gap) $\ol g_{n}^+$ and the lower rim (gap) $\ol g_{n}^{\ -}$, i.e.,
\[
g_n^c=\ol g_{n}^+\cup \ol g_{n}^-,\ \ where \ \ g_{n}^\pm =g_n\pm i0;
\qq and \  \ if \ z\in g_n \Rightarrow z\pm i0\in g_n^\pm.
\]

\begin{figure}
\tiny
\unitlength=1mm
\special{em:linewidth 0.4pt}
\linethickness{0.4pt}
\begin{picture}(120.67,34.33)
\put(20.33,21.33){\line(1,0){100.33}}
\put(70.33,10.00){\line(0,1){24.33}}
\put(69.00,19.00){\makebox(0,0)[cc]{$0$}}
\put(120.33,19.00){\makebox(0,0)[cc]{$\Re z$}}
\put(67.00,33.67){\makebox(0,0)[cc]{$\Im z$}}
\put(81.33,21.33){\linethickness{2.0pt}\line(1,0){9.67}}
\put(100.33,21.33){\linethickness{2.0pt}\line(1,0){4.67}}
\put(116.67,21.33){\linethickness{2.0pt}\line(1,0){2.67}}
\put(60.00,21.33){\linethickness{2.0pt}\line(-1,0){9.33}}
\put(40.00,21.33){\linethickness{2.0pt}\line(-1,0){4.67}}
\put(24.33,21.33){\linethickness{2.0pt}\line(-1,0){2.33}}
\put(81.67,24.00){\makebox(0,0)[cc]{$e_1^-$}}
\put(91.00,24.00){\makebox(0,0)[cc]{$e_1^+$}}
\put(100.33,24.00){\makebox(0,0)[cc]{$e_2^-$}}
\put(105.00,24.00){\makebox(0,0)[cc]{$e_2^+$}}
\put(115.33,24.00){\makebox(0,0)[cc]{$e_3^-$}}
\put(120.00,24.00){\makebox(0,0)[cc]{$e_3^+$}}
\put(59.33,24.00){\makebox(0,0)[cc]{$-e_1^-$}}
\put(50.67,24.00){\makebox(0,0)[cc]{$-e_1^+$}}
\put(40.33,24.00){\makebox(0,0)[cc]{$-e_2^-$}}
\put(34.67,24.00){\makebox(0,0)[cc]{$-e_2^+$}}
\put(26.00,24.00){\makebox(0,0)[cc]{$-e_3^-$}}
\put(19.50,24.00){\makebox(0,0)[cc]{$-e_3^+$}}
\end{picture}
\caption{The cut domain $\cZ=\C\sm \cup \ol g_n$ and the slits $g_n=(e_n^-,e_n^-)$ in the $z$-plane.}
\lb{z}
\end{figure}

In order to construct the  Riemann surface $\cM$ we take the cut domain $\cZ=\C\sm \cup \ol g_n$ and identify (we glue) the upper rim $g_{n}^+$ of the slit $g_n^c$ with the upper rim $g_{-n}^+$  of the slit $g_n^c$ and correspondingly the lower rim $g_{n}^-$ of the slit $g_{-n}^c$ with the  lower  rim $g_{-n}^-$  of the slit $g_{-n}^c$ for all nontrivial gaps. The mapping $z=\sqrt \l$ from $\L$ onto $\cM$ is one-to-one and onto. The gap $\g_n^{(1)}\ss \L_1$ is mapped onto $g_n^+\ss \cM_1$ and the gap $\g_n^{(2)}\ss \L_2$ is mapped onto $g_n^-\ss \cM_2$. From a physical point of view, the upper rim  $g_{n}^+$ is a physical gap and the lower rim  $g_{n}^-$ is a non physical gap.
Moreover, $\cM\cap\C _+=\cZ\cap\C _+$ plus all physical gaps $g_{n}^+$ is a so-called physical "sheet" $\cM_1$
and $\cM\cap\C _-=\cZ\cap\C _-$ plus all non physical gaps $g_{n}^-$ is a so-called non physical "sheet" $\cM_2$. The set (the spectrum) $\R\sm \cup g_n$ joints the first $\cM_1$ and second sheets $\cM_2$.

\begin{figure}
\tiny
\unitlength=1mm
\special{em:linewidth 0.4pt}
\linethickness{0.4pt}
\begin{picture}(120.67,34.33)
\put(20.33,20.00){\line(1,0){102.33}}
\put(71.00,7.00){\line(0,1){27.00}}
\put(70.00,18.67){\makebox(0,0)[cc]{$0$}}
\put(124.00,18.00){\makebox(0,0)[cc]{$\Re k$}}
\put(67.00,33.67){\makebox(0,0)[cc]{$\Im k$}}
\put(87.00,15.00){\linethickness{2.0pt}\line(0,1){10.}}
\put(103.00,17.00){\linethickness{2.0pt}\line(0,1){6.}}
\put(119.00,18.00){\linethickness{2.0pt}\line(0,1){4.}}
\put(56.00,15.00){\linethickness{2.0pt}\line(0,1){10.}}
\put(39.00,17.00){\linethickness{2.0pt}\line(0,1){6.}}
\put(23.00,18.00){\linethickness{2.0pt}\line(0,1){4.}}
\put(85.50,18.50){\makebox(0,0)[cc]{$\pi$}}
\put(54.00,18.50){\makebox(0,0)[cc]{$-\pi$}}
\put(101.00,18.50){\makebox(0,0)[cc]{$2\pi$}}
\put(36.00,18.50){\makebox(0,0)[cc]{$-2\pi$}}
\put(117.00,18.50){\makebox(0,0)[cc]{$3\pi$}}
\put(20.00,18.50){\makebox(0,0)[cc]{$-3\pi$}}
\put(87.00,26.00){\makebox(0,0)[cc]{$\pi+ih_1$}}
\put(56.00,26.00){\makebox(0,0)[cc]{$-\pi+ih_1$}}
\put(103.00,24.00){\makebox(0,0)[cc]{$2\pi+ih_2$}}
\put(39.00,24.00){\makebox(0,0)[cc]{$-2\pi+ih_2$}}
\put(119.00,23.00){\makebox(0,0)[cc]{$3\pi+ih_3$}}
\put(23.00,23.00){\makebox(0,0)[cc]{$-3\pi+ih_3$}}
\end{picture}
\caption{The domain $\cK=\C\sm \cup \G_n$, where the slit $\G_n=(\pi n-ih_n,\pi n+ih_n)$}
\lb{k}
\end{figure}

We introduce the quasimomentum $k(\cdot )$ for $H_0$ by $k(z)=\arccos \D(z),\ z \in \cZ$. The function
$k(z)$ is analytic in $z\in\cZ$ and satisfies:
\[
\lb{pk}
(i)\qq k(z)=z+O(1/z)\qq  as \ \ |z|\to \iy, \qq  \qq (ii)\qq  \Re k(z\pm i0) |_{[e_n^-,e_n^+]}=\pi n,\qq \ n\in \Z,
\]
and $\pm \Im k(z)>0$ for any $z\in \C_\pm$, see \cite{MO}, \cite{KK}. The function $k(\cdot)$ is analytic
on $\cM$ and satisfies $\sin k(z)=(1-\D^2(z))^{1\/2}, z\in \cM$. Moreover, the quasimomentum $k(\cdot)$ is a conformal mapping from $\cZ$ onto the quasimomentum domain $\cK=\C\sm \cup \G_n$, see  Fig. 2  and 3. Here $\G_n=(\pi n-ih_n,\pi n+ih_n)$ is a vertical slit with the height $h_n\ge 0, h_0=0$.
The height $h_n$ is defined by the equation $\cosh h_n=(-1)^n\D(e_n)\ge 1$, where $e_n\in [e_n^-,e_n^+]$ and $\D'(e_n)=0$. The function $k(\cdot)$ maps the slit $g_n^c$ on the slit $\G_n,$ and $k(-z)=-k(z)$ for all $z\in \cZ$.

In order to describe the spectral properties of the operator $H_0$ we need the properties of $\vt, \vp$.
Recall that  $\vt,\vp$ are the solutions of the equation $-y''+py=z^2y$ with the conditions $\vp'(0,z)=\vt(0,z)=1$ and $\vp(0,z)=\vt'(0,z)=0$. The solutions $\vt,\vp$ satisfy the equations
\begin{multline}
\lb{fs}
\vt(x,z)=\cos zx+\int_0^x{\sin z(x-s)\/z}p(s)\vt(s,z)ds,\qqq  \\
\vp(x,z)={\sin zx\/z}+\int_0^x{\sin z(x-s)\/z}p(s)\vp(s,z)ds.
\end{multline}
For each $x\in \R$ the functions  $\vt(x,z),  \vp(x,z)$  are entire in $z\in\C$ and satisfy
\begin{multline}
\lb{efs1}
\max \rt\{|z|_1|\vp(x,z)|, \ |\vp'(x,z)| , |\vt(x,z)|,
{1\/|z|_1}|\vt'(x,z)|    \rt\} \le X=e^{|\Im z|x+\|p\|_x},\\
\rt|\vp(x,z)-{\sin zx\/z}\rt|\le {X\/|z|^2}\|p\|_x,
\qq |\vt(x,z)-{\cos zx}|\le {X\/|z|}\|p\|_x,\qqq |z|_1=\max\{1, |z|\},
\end{multline}
where   $\|p\|_t=\int_0^t|p(s)|ds$ and $(x,z)\in \R\ts \C$, see  [PT]. These estimates yield
\[
\lb{asb} \b(z)={\vp'(1,z)-\vt(1,z)\/2}=\int_0^1{\sin z(2x-1)\/z}p(x)dx+{O(e^{|\Im z|})\/z^2}\qqq \as \qqq |z|\to \iy.
\]
Moreover, if $z=\pi n+O(1/n)$, then we obtain
\[
\lb{abn}
\b(z)=(-1)^n{p_{sn}+O(n^{-1})\/2\pi n},\qq
p_{sn}=\int_0^1p(x)\sin 2\pi nxdx.
\]

The Floquet solutions $\p^{\pm}(x,z), z \in \cZ$ of the equation $-y''+py=z^2y$ are given by
\[
\lb{3}
\p^\pm(x,z)=\vt(x,z)+m_\pm(z)\vp(x,z),\qqq \ m_\pm={\b\pm i\sin k\/ \vp(1,\cdot)},
\]
where $\vp(1,z)\p^+(\cdot,z)\in L^2(\R_+)$ for all $z\in\C_+\cup\cup_{} g_n$.
If $p=0$, then $k=z$ and $\p^\pm(x,z)=e^{\pm izx}$.

Let $\cD_r(z_0)=\{|z-z_0|<r\}$ be a disk for some $r>0, z_0\in \L$. It is well known that if $g_n=\es$ for some $n\in \Z$, then the functions $\sin k(\cdot)$ and $ m_\pm$ {\bf are analytic in some disk}
$\cD_\ve(\m_n)\ss\cZ, \ve>0$ and the functions $\sin k(z)$ and $\vp(1,z)$ have the simple zero at $\m_n$, see \cite{F1}. Moreover, $m_\pm$ satisfies
\[
\lb{Tm-2}
m_\pm (\m_n)={\b'(\m_n)\pm i(-1)^nk'(\m_n)\/\pa_z\vp(1,\m_n)},
\qq \Im m_\pm (\m_n)\ne 0.
\]
Furthermore, $\Im m^+ (z)>0$ for all $(z,n)\in (z_{n-1}^+,z_{n}^-)\ts \N$ and the asymptotics hold true:
\begin{multline}
\lb{Tm-1}
m_\pm (z)=\pm iz+O(1) \qq as \qq |z|\to \iy, \qq z\in \cZ_\ve
\\
\where \qqq \cZ_\ve =\{z\in \cZ: \dist \{z,g_n\}>\ve, g_n\ne \es,  n\in \Z\},\ \ve>0.
\end{multline}
The function $\sin k$ and each function $\vp(1,\cdot)\p^{\pm}(x,\cdot), x\in \R$ are analytic on the Riemann surface $\cM$. Recall that the Floquet solutions $\p_\pm(x,z), (x,z)\in \R\ts \cM$ satisfy (see \cite{T})
\[
\lb{f1}
\p_\pm(0,z)=1, \qq \p^\pm(0,z)'=m_\pm(z),
\qq \p^\pm(1,z)=e^{\pm ik(z)}, \qq  \p^\pm(1,z)'=e^{\pm ik(z)}m_\pm(z),
\]
\[
\lb{f2}
\p^\pm(x,z)=e^{\pm ik(z)x}(1+O(1/z)) \qq  .
\]
as $|z|\to \iy \ z\in \cZ_\ve$, uniformly in $x\in \R$. Below we need the simple identities
\[
\lb{LD0}
\b^2+1-\D^2=1-\vp'(1,\cdot)\vt(1,\cdot)= -\vp(1,\cdot)\vt'(1,\cdot).
\]

Introduce the fundamental solutions $\P^\pm (x,z)$ of the equation
\[
\lb{bcf}
-{\P^\pm}''+(p+q)\P^\pm=z^2 \P^\pm, x\ge 0,\qq
\P^\pm(x,z)=\p^\pm(x,z),\ \ x\ge t , \qq z\in\cZ\sm \{0\}.
\]
Each function $\vp(1,z)\P^\pm(x,z),x\ge 0$ is analytic  in $\cM$, since each $\vp(1,z)\p^{\pm}(x,z), x\ge 0$
is  analytic in $\cM$. We define {\bf the modified Jost function} $\P_0^\pm=\P^\pm(0,z)$,
which is meromorphic in $\cM$ and has branch points $e_n^\pm, g_n\ne \es$.
The kernel of the resolvent $R=(H-z^2)^{-1},  z\in \C_+,$ has the form
\[
\lb{R}
R(x,x',z)={\F (x,z )\P^+(x',z)\/\P_0^+(z)},\ \ \ x<x',\ \ and \ \
R(x,x',z )=R(x',x,z ),\ x>x'.
\]
Recall that $\F (x,z)$ is the solution of the equation
$-\F''+(p+q)\F=z^2 \F, x\ge 0,\ \F (0,z)=0, \ \F'(0,z)=1, \ z\in\C.$
Each function $R(x,x',z), x,x'\in \R$ is meromorphic in $\cM$. Remark that if
$z_0\in g_n^\pm\sm \{\m_n\pm i0\}$ and $\P_0^+(z_0)\ne 0$ for some $n$, then the resolvent of $H$ is analytic at $z_0$.
The function $\P_0^+(z)$ has finite number of simple zeros
 on each  $g_n^+, n\ne 0$ and on $i\R_+$ (no zeros on $\C_+\sm i\R_+$), where the squared zero is an eigenvalue.
A pole of $\mR(x,z)=\P^+(x,z)/\P_0^+(z)$ on $g_n^+$ is called a bound state.
Of course, $z^2$ is really the energy, but since
the momentum $z$ is the natural parameter, we will abuse the terminology.
 Moreover, $\P_0^+(z)$ has infinite number of zeros in  $\C_-$, see \er{T2-3}.
 We rewrite Definition $\L$ about the resonances on the Riemann surface $\L$
 in the equivalent form in terms of the resonances on the Riemann surface $\cM$.

\no {\bf Definition $\cM$. } {\it
1) A point $\z\in \ol\C_-\cap \cM, \z\ne 0$ is a resonance, if the function
$\mR(x,z)$ has a pole at $\z$  for almost all $x>0$.

\no 2) A point  $\z=e_n^\pm , n\ne 0$ (or $\z=0$) is a virtual state, if  the function $\mR(x,\z+z^2)$ (or $\mR(x,z)$) has a pole at $\z$  for almost all $x>0$.

\no 3) A point $\z\in\cM$ is a state, if it is a bound state or a resonance or a virtual state.
If $\z\in g_n^-, n\ne 0$ or $\z\in g_0^-=i\R_-$, then we call $\z$ an antibound state.
}

Let  $\s_{bs}(H)$ (or $\s_{rs}(H)$ or $\s_{vs}(H)$ ) be the set of all  bound states
in the momentum variable $z=\sqrt \l\in \cM$ (or resonances or virtual states)
of $H$ and let $\s_{st}(H)=\s_{vs}(H)\cup \s_{rs}(H)\cup\s_{bs}(H)$.

The kernel of the resolvent $R_0(z)=(H_0-z^2)^{-1}, z\in \C_+$ has the form
\[
\lb{R0}
R_0(x,x',z)=\vp(x,z)\p^+(x',z),\qq x<x',\ \
{\rm and} \qq R_0(x,x',z)=R_0(x',x,z),\ x>x'.
\]
Consider the states $z_n^0=\sqrt{\l_n^0}\in g_n^c$ of $H_0$.
Due to  \er{3}, the function $\P_0^+=\p_0^+(0,\cdot)=1$ and $\mR(x,\cdot)=\vt(x,\cdot)+m_+\vp(x,\cdot)$.
Recall that $\vp(1,\cdot)m_+$ and $\sin k(z)$ are analytic in $\cM$.
Thus the resolvent $R_0(z)$ has singularities only at $\m_n\pm i0$, where $g_n\ne \es$,
and in order to describe the states of $H_0$ we need to study $m_+$ on $g_n^c$ only.
Let $\mA(z_0), z_0\in \cM$ be the set of analytic functions in some disk $\cD_r(z_0)=\{|z-z_0|<r\}, r>0$.
We need the following result (see \cite{Zh3})

\begin{lemma}
\lb{Tm}
All states of $H$ are given by $\m_n\pm i0\in g_n^c, n\ne 0$, where $g_n\ne \es$.
Let the momentum gap $g_n=(e_n^-,e_n^+)\ne \es$ for some $n\ge 1$. Then

\no i)  $z_n^0=\m_n+i0\in g_n^+$ is a bound state of $H_0$ iff one condition from (1)-(3) holds true
\begin{multline}
\lb{Tm-31}
(1) \qq  m_-\in \mA(\m_n+i0),\\
(2)\ \ \b(\m_n)=i\sin k(\m_n+i0)= -(-1)^n\sinh h_{sn},  \qq  k(\m_n+i0)=\pi n+ih_{sn}\qq h_{sn}>0  \hspace{1.2cm} \\
(3)\qqq  m_+(z_n^0+z)={c_n+O(z)\/z} \qq as \ z\to 0, \ z\in \C_+,\
c_n={-2\sinh |h_{sn}|\/(-1)^n\pa_z\vp(1,\m_n)}<0.\qq
\end{multline}
\no ii)  $z_n^0=\m_n-i0\in g_n^-$ is an antibound state of $H_0$ iff
one condition from (1)-(3) holds true
\begin{multline}
\lb{Tm-32}
(1) \qqq   m_-\in \mA(\m_n-i0),\\
(2)\qqq
\b(\m_n)=i\sin k(\m_n-i0)= -(-1)^n\sinh h_{sn},  \qq k(\m_n-i0)=\pi n+ih_{sn},\qq h_{sn}<0,    \hspace{0.3cm} \\
(3)\qqq  \qq m_+(z_n^0+z)={-c_n+O(z)\/z} \qq as \ z\to 0, \ \ z\in \C_-.  \hspace{4.5cm}
\end{multline}

\no iii) $z_n^0=\m_n$ is a virtual state of $H_0$ iff one condition from (1)-(2) holds true
\begin{multline}
\lb{Tm-33}
(1) \qqq z_n^0=\m_n=e_n^-\qqq  or \qqq z_n^0=\m_n=e_n^+,\\
(2) \qqq m_+(z_n^0+z)={c_n^0+O(z)\/\sqrt z}\qq as \ z\to 0, \ z\in \C_+,\qq
c_n^0\ne 0. \hspace{3.7cm}
\end{multline}
\end{lemma}

These simple facts are well known in the inverse spectral theory, see \cite{N-Z}, \cite{MO} or \cite{K5}.
Remark that detail analysis of $H_0$ was done in \cite{Zh3}.


If $\m_n\in g_n\ne \es$, then the function $m_+$ has a pole at $z_n^0=\m_n+i0\in g_n^+$ (a bound state) or at $z_n^0=\m_n-i0\in g_n^-$ (an antibound state). If $\m_n=e_n^+$ (or $\m_n=e_n^-$), then $z_n^0=\m_n$ is a virtual state.
Note that if some $g_n=\es, n\ne 0$, then each $\p^\pm(x,\cdot)\in \mA(\m_n), x\ge 0$.
Moreover, the  resolvent $R_0(z)$ has a pole at $z_0$ iff the function $m_+(\cdot)$ has a pole at $z_0$.


 The following asymptotics from \cite{MO} hold true as $n\to \iy$:
\[
\lb{sde}
\m_n=\pi n+\ve_n(p_{0}-p_{cn}+O(\ve_n)),\qqq p_{cn}=\int_0^1p(x)\cos 2\pi nxdx,
\qqq
\qq \ve_n={1\/2\pi n},
\]
\[
\lb{anc}
h_{sn}=-\ve_n(p_{sn}+O(\ve_n)),
\]
\[
\lb{ape}
e_n^\pm=\pi n+\ve_n(p_0\pm |p_n|+O(\ve_n)), \qq \qq
p_n=\int_0^1p(x)e^{-i2\pi nx}dx=p_{cn}-ip_{sn}.
\]

Let $\vp(x,z,\t), \ (z,\t)\in \C\ts \R$   be the solutions of the equation
\[
\lb{x+t}
-\vp''+p(x+\t)\vp=z^2 \vp, \qq \ \vp(0,z,\t)=0,\qq \vp'(0,z,\t)=1.
\]
Let  $ y_1,  y_2$ be the solutions of the equations $-y''+(p+q)y=z^2y, z\in \C$ and satisfying
\[
\lb{wtc}
y_2'(t,z)=y_1(t,z)=1,\qqq y_2(t,z)=y_1'(t,z)=0.
\]
Thus, they satisfy the integral equation
\begin{multline}
\lb{efy}
y_1(x,z)=\cos z(x-t)-\int_x^t{\sin z(x-\t)\/z}(p(\t)+q(\t))y_1(\t,z)d\t,\qqq  \\
y_2(x,z)={\sin z(x-t)\/z}-\int_x^t{\sin z(x-\t)\/z}(p(\t)+q(\t))y_2(\t,z)d\t.
\end{multline}
For each $x\in \R$ the functions  $ y_1(x,z),  y_2(x,z)$  are entire in $z\in\C$ and satisfy
\begin{multline}
\lb{efs}
\max \{||z|_1 y_2(x,z)|, \ | y_2'(x,z)| , | y_1(x,z)|,
{1\/|z|_1}| y_1'(x,z)|    \} \le X_1=e^{|\Im z||t-x|+\int_x^t|p+q|ds},\\
| y_1(x,z)-\cos z(x-t)|\le {X_1\/|z|}\|q\|_t,\qq
\rt| y_2(x,z)-{\sin z(x-t)\/z}\rt|\le {X_1\/|z|^2}\|q\|_t,
\end{multline}
and recall that $|z|_1=\max\{1, |z|\}$ and  $\|p\|_t=\int_0^t|p(s)|ds$.

The equation $-y''+(p-z^2)y=f, y(0)=y'(0)=0$ has an unique solution given by
$
y=\int_0^x\vp(x-\t,z,\t)f(\t)d\t.
$
Hence the solutions $\F$ and $\P_\pm$  of the  $-y''+(p+q)y=z^2y$ satisfy
\[
\lb{eF}
\F(x,z)=\vp(x,z)+\int_0^x\vp(x-s,z,s)q(s)\F (s,z)ds,
\]
\[
\lb{ep}
\P^\pm (x,z)=\p^\pm(x,z)-\int_x^t\vp(x-s,z,s)q(s)\P^\pm(s,z)ds.
\]
Below we need the well known fact for scattering theory
\[
\lb{DP}
\P^+(0,z)=D(z^2)=\det (I+q(H_0-z^2)^{-1}),\qqq z\in \cM.
\]
It is similar to the case $p=0$, see \cite{J}.
The case on the real with $p\ne \const$ was considered in \cite{F4}.
The functions $\P^\pm, m_\pm, ,...$ are meromorphic in $\cM$ and real on $i\R$. Then the following identities hold true:
\[
\P^\pm(-z)=\ol \P^\pm(\ol z), \qq m_\pm(-z)=\ol m_\pm(\ol z), \qq ...,\qq  z\in \cZ.
\]

\begin{lemma}
\lb{T21}
i) The following identities and asymptotics  hold true:
\[
\lb{T21-1}
\P^\pm =\p^{\pm}(t,\cdot)y_1+\dot\p^{\pm}(t,\cdot)y_2,\qqq \where \qq \dot u=\pa_t u,
\]
\[
\lb{T21-2}
\P^\pm(0,z)=1+\int_0^t\vp(x,z)q(x)\P^\pm(x,z)dx,
\]
\[
\lb{T21-3}
\P^\pm(x,z)=e^{\pm ik(z)x}(1+e^{\pm (t-x)(|v|-v)}O(1/z)),\qq
v=\Im z,\qq    \qq
\]
as $|z|\to \iy, z\in \cZ_\ve, \ve>0$ uniformly in $x\in [0,t]$.
Moreover, \er{T1-1} hold true.

\no ii) The function $\P^\pm(0,\cdot)$ has
exponential type $2t$ in the half plane $\C_\mp$.

\end{lemma}
\no{\bf Proof.} i)  Using \er{wtc}, \er{bcf} we obtain \er{T21-1}.

The identity $\vp(x,\cdot,\t)=\vt(\t,z)\vp(x+\t,z)-\vp(\t,z)\vt(x+\t,z)$ gives
$\vp(-x,\cdot,x)=-\vp(x,z)$ and \er{ep} yield \er{T21-2}.

Substituting \er{f2} into \er{ep}  we obtain \er{T21-3}.
In particular, substitution of \er{T21-3}, \er{efs} into \er{T21-2} yields
\er{T1-1}, since $D(z^2)=\P^+(0,z)$.

ii) We give the proof for the case $\P^+(0,z)$, the proof for $\P^-(0,z)$ is
similar. Due to \er{T21-3}, $\P^+(0,z)$ has exponential type $\le 2t$ in the half plane $\C_-$.
The decompositions $f(x,z)\ev e^{-ixk(z)}\P^+(x,z)=1+\ve f_1(x,z)$
and $\vp(x,z)e^{ixk(z)}\ev \ve (e^{i2xz}-1 +\ve \e(x,z)), \ \ve={1\/2iz}$ give
\begin{multline}
\lb{esff}
\P^+(0,z)-1=\int_0^t\vp(x,z)e^{ixk}q(x)f(x,z)dx\\
=
\ve \int_0^te^{i2xz}q(x)f(x,z)(1+\ve e^{-i2xz}\e(x,z))dx-\ve \int_0^tq(x)f(x,z)dx=\ve K(z)-\ve \int_0^tq(x)dx,\\
K=\ve \int_0^te^{i2xz}q(x)(1+G(x,z))dx,
\qqq G=\ve f_1+\ve e^{-i2xz}(\e f-f_1), \qq  \  z\in \cZ_\ve.
\end{multline}

Asymptotics \er{efs1}, \er{T21-3} and $k(z)=z+O(1/z)$ as $z\to \iy$ (see \cite{KK}) yield
\[
\lb{esff1}
\e(x,z)=
e^{2x|\Im z|}O(1),\qq
\qqq f_1(x,z)=e^{2(t-x)|\Im z|}O(1)\qq as \ |z|\to \iy, z\in \cZ_\ve.
\]

We need the following variant of the Paley Wiener Theorem from
\cite{Fr}:

\no {\it let $h\in \cQ_t$ and let each $F(x,z), x\in [0,t]$ be analytic for $z\in\C_-$ and $F\in L^2((0,t)dx,\R dz)$. Then $\int_0^te^{2izx}h(x)(1+F(x,z))dx$ has exponential type at least $2t$ in $\C_-$.}

We can not apply this result to the function  $K(z), z\in\C_-$, since $\p^+(x,z)$ has a singularity at $z_n^0$ if
$g_n\ne \es$.  But we can use this result for the function  $K(z-i), z\in\C_-$,
since \er{esff}, \er{esff1} imply $\sup _{x\in [0,1]}|G(x,-i+\t)|=O(1/\t)$ as $\t\to \pm\iy$.
Then the function $\P^+(0,z)$ has exponential type $2t$ in the half plane $\C_-$.
 \BBox

\section {Spectral properties of $H$}
\setcounter{equation}{0}

Recall that an entire function $f(z)$ is said to be of $exponential$
$ type$  if there is a constant $A$ such that  $|f(z)|\le $ const.
$e^{A|z|}$ everywhere (see [Koo]). The infimum over the set of $A$
for which such an inequality holds is called the type of $f$. {\it
The function $f$ is said to belong to the Cartwright class $Cart_\o$
if $f(z)$ is entire, of exponential type, $\o_\pm(f)=\o>0$, where $
\o_{\pm}(f)=\lim \sup_{y\to \iy} {\log |f(\pm iy)|\/y}$ and $\int
_{\R}{\log ^+|f(x)|dx\/ 1+x^2}<\iy$.}

Let for shortness
$$ \vp_1=\vp(1,z ),\ \ \ \vp_1'=\vp'(1,z ), \ \ \ \vt_1=\vt(1,z ),....,
  \F_1=\F(1,z),\ \ \ \F_1'=\F'(1,z).
$$

\no {\bf Define the functions}
\[
F(z)=\vp(1,z)\P^-(0,z)\P^+(0,z), \qq z\in \cZ,
\]
which is real on $\R$,  since $\P^-(0,z)=\ol\P^+(0,\ol z)$ for all $z\in \cZ$,
see also \er{T22-1}.

\begin{lemma} \lb{T22}
i) The following identities and estimates hold true:
\[
\lb{T22-1}
F=\vp(1,\cdot,t)y_1^2(0,\cdot)+\dot\vp(1,\cdot,t)y_1(0,\cdot)y_2(0,\cdot)-\vt'(1,\cdot,t)y_2^2(0,\cdot)\in Cart_{1+2t},
\]
\[
\lb{T22-2}
\vp_1\dot\p_t^+\dot\p_t^-=-\vt'(1,\cdot,t),\qqq
\vp_1(\dot\p_t^+\p_t^-+\p_t^+\dot\p_t^-)=\dot\vp(1,\cdot,t)=
\vp'(1,\cdot,t)-\vt(1,\cdot,t),
\]
\[
\lb{T22-3}
|F(z)-{\sin z\/z}|\le {C_F e^{(1+2t)|\Im z|}\/|z|^2},
\qq C_F=3(\|p\|_1+\|p+q\|_t)e^{2\|p+q\|_t+\|p\|_1}.
\]

\no ii) The set of zeros of $F$ is
symmetric with respect to the real line and the imaginary line.
In each disk $\{z: |z-\pi n|<{\pi\/ 4}\}, |n|\ge 1+4C_Fe^{t{\pi\/2}}$ there exists exactly one simple real zero $z_n$ of $F$ and $F$
has not zeros in the domain $\mD_F\cap\C_-$, where $\mD_F=\{z\in \C: 4C_Fe^{2|\Im z|}<|z|\}$.

\no iii) For all $z \in \cZ$ the following identity holds true:
 \[
\lb{T31-1}
\P_0^\pm(z)=e^{\pm ik(z)n_t}w_\pm(z),\ \ \  w_\pm(z)=\F'(n_t,z)-m_\pm(z)\F(n_t,z), \qqq n_t=\inf_{n\in \N, n\ge t} n.
\]

\end{lemma}
\no{\bf Proof.} i)
The function $\vp(1,z,t)$ for all $(t,z)\in R\ts \C$ satisfies the following identity
\[
\lb{ipv}
\vp(1,\cdot,t)=-\vt_1'\vp_t^2+\vp_1\vt_t^2+
2\b\vp_t\vt_t=\vp_1\p_t^+\p_t^-,
\]
see \cite{Tr}.
Recall that $\vp(1,z,t), \vt(1,z,t)$ are define in \er{x+t}.   Using \er{ipv}  we obtain
\begin{multline}
\lb{2is}
\dot \vp(1,\cdot,t)=\vp_1(\dot \p_t^+\p_t^-+\p_t^+\dot\p_t^-),\\
\ddot \vp(1,\cdot,t)=\vp_1(\ddot \p_t^+\p_t^-+\p_t^+\ddot\p_t^-
+2\dot \p_t^+\dot \p_t^-)=2\vp_1(p(t)-z^2)\p_t^+\p_t^-+
2\vp_1\dot \p_t^+\dot \p_t^-.
\end{multline}
Identity \er{T21-1} gives
\[
\lb{e12}
F=\vp_1(\p_t^+\p_t^-y_1^2(0,\cdot)+\dot\p_t^+\dot\p_t^-y_2^2(0,\cdot)+
(\p_t^+\dot\p_t^-+\dot\p_t^+\p_t^-)y_1(0,\cdot)y_2(0,\cdot).
\]
Then using the following identities from \cite{IM}
\[
\lb{im1}
\dot \vp(1,z,t)=\vp'(1,z,t)-\vt(1,z,t),\qq
\ddot \vp(1,z,t)=2(p(t)-z^2)\vp(1,z,t)-2\vt'(1,z,t),
\]
and  \er{ipv}, \er{e12} we obtain \er{T22-1},\er{T22-2}, since Lemma \ref{T21}, ii) and \er{efs1}
yields $F\in Cart_{1+2t}$.

We will show \er{T22-3}. We have
\[
y_1(0,\cdot)=\cos t z+\wt y_1,\qqq \wt y_1=\int_0^t
{\sin z s\/z}(p(s)+q(s))y_1(s,\cdot)ds,
\]
\[
y_2(0,\cdot)=-{\sin t z\/z}+\wt y_2,\qqq \wt y_2=\int_0^t
{\sin z s\/z}(p(s)+q(s))y_2(s,\cdot)ds,
\]
\[
\vt(1,t)=\cos z+\vt_{1t},\qq \vt_{1t}=\int_0^1
{\sin z(1-s)\/z}p(s+\t)\vt(s,\t)ds,
\]
\[
\vt'(1,t)=-z\sin z+\vt_{1t}',\qq \vt_{1t}'=\int_0^1
\cos z(1-s)p(s+t)\vt(s,t)ds,
\]
\[
\vp(1,t)={\sin z\/z}+\vp_{1t},\qq \vp_{1t}=\int_0^1
{\sin z(1-s)\/z}p(s+t)\vp(s,t)ds,
\]
Then \er{T22-1} imply
$$
F=(\cos tz+\wt y_1)^2({\sin z\/z}+\vp_{1t})+({\sin tz\/z}-\wt y_2)^2(z\sin z-\vt_{1t}')
+(\cos tz+\wt y_1)(-{\sin tz\/z}+\wt y_2)\dot \vp(1,z,t)
$$
$$={\sin t z\/z}+f_1+f_2+f_3,
$$
where
$$
 f_1=y_1(0,\cdot)^2\vp_{1t}+{\sin tz\/z}(\cos tz+y_1(0,\cdot) )\wt y_1,
$$$$
f_2=-y_2(0,\cdot)^2\vt_{1t}'+z\sin z(y_{2}(0,\cdot)-{\sin tz\/z})\wt y_2,
\qqq  f_3=y_1(0,\cdot)y_2(0,\cdot)\dot \vp(1,z,t),
$$
$$
\qq
 |f_3|\le {C_t\/|z|^2}\|p\|_1,\qq
|f_j|\le {C_t\/|z|^2}(\|p\|_1+2\|p+q\|_t),\qq j=1,2,
$$
which yields \er{T22-3},
where $\|q\|_t=\int_0^t|q(x)|dx$ and $C_t=e^{(2t+1)|\Im z|+2\|p+q\|_t+\|p\|_1}$.

ii) Using \er{T22-3} we obtain for $|n|\ge 1+4C_Fe^{t{\pi\/2}}$
$$
|F(z)-{\sin z\/z}|\le {C_F\/|z|^2}e^{|\Im z|+t{\pi\/2}|}
\le {4C_F\/|z|}e^{t{\pi\/2}|}{|\sin z|\/|z|}<{|\sin z|\/|z|}
 \qq all \ |z-\pi n|={\pi\/4},
$$
since $e^{|{\Im}z|}\le 4|\sin z|$ for all
$|z-\pi n|\ge \pi /4, n\in \Z$, (see [PT]).
Hence, by Rouche's theorem, $F$ has as many roots, counted with
multiplicities, as $\sin z$ in each disk $\cD_{\pi\/4}(\pi n)$. Since $\sin z$ has only the roots
$\pi n, n\ge 1,$ and i) of the lemma follows.
This zero in $\cD_{\pi\/4}(\pi n)$ is real since $F$ is real
on the real line.

Using \er{T22-3} and $e^{|{\Im}z|}\le 4|\sin z|$ for all
$|z-\pi n|\ge \pi /4, n\in \Z$, we obtain
$$
|F(z)|\ge |{\sin z\/z}|-\rt|F(z)-{\sin z\/z}\rt|\ge {e^{|\Im z|}\/4|z|^2}
\rt(|z|-4C_F e^{2t|\Im z|}   \rt)>0,
$$
for all $z\in \mD_1=\{z\in \mD_F: |z-\pi n|\ge \pi /4, n\in \Z \}$.
This yields $|F|>0$ in $\mD_1$. But the function $F$ has
exactly one real zero $z_n$ in $\cD_{\pi\/4}(\pi n), n\ge n_0$.
Then $F$ has not zeros in the domain $\mD_F$.
The function  $F$ is real on the real line and the imaginary line.
Hence  the set of zeros of $F$ is symmetric with respect to the real line and the imaginary line.

iii) Using $\P^\pm(0,z)=\{\P^\pm(x,z ),\F (x,z)\},  \ z \in \cZ$ at $x=n_t$,
and  \er{f1} we obtain  \er{T31-1}, where   $\{y,u\}=yu'-y'u$ is the Wronskian.
\BBox

Let  $\wt\vt, \wt\vp$ be the solutions of the equations
$-y''+(p+q)y=z^2y, z\in \C$ and satisfying
$$
\wt\vt(x,z)=\vt(x,z),\qqq \wt\vp(x,z)=\vp(x,z),\qq \all \qq x\ge t.
$$
Hence the solutions $\wt\vt, \wt\vp$  satisfy the equations
\begin{multline}
\lb{eqfwt}
\wt\vt(x,z)=\vt(x,z)-\int_x^t\vp(x-s,z,s)q(s)\wt\vt(s,z)ds,\\ \wt\vp(x,z)=\vp(x,z)-\int_x^t\vp(x-s,z,s)q(s)\wt\vp(s,z)ds.
\end{multline}
For each $x\ge 0$ the functions $\wt\vt(x,z), \wt\vp(x,z)$ are entire and real for $z^2\in \R$.
The identities \er{eqfwt} and \er{bcf} give
\[
\lb{Ptp}
\P^\pm(x,z)=\wt\vt(x,z)+m_\pm(z)\wt\vp(x,z),\qqq \all \qq (x,z)\in \R_+\ts \cZ.
\]

Recall that $g_n^c={\ol g_n^-}\cup {\ol g_n^+}$ and we define the sets
\[
\mZ=i\R\cup \C_-\cup \cup_{n\in \Z}g_n^c,\qqq \qq  \mZ_0=\mZ\sm \{0, e_n^\pm,\m_n\pm i0, g_n\ne \es, n\in \Z\}.
\]

\begin{lemma} \lb{T311}
i) If $g_n=(e_n^-,e_n^+)= \es$ for some $n\ne 0$, then each $\P^\pm(x,\cdot )\in \mA(\m_n), x\ge 0$ and
$\Im \P^\pm(0,\m_n)\ne 0$. Moreover, $\m_n=e_n^\pm$ is a simple zero of $F$
and $\m_n\notin\s_{st}(H)$.

\no ii)  $\P^\pm(0,z)\ne 0$ for all $z\in (e_{n-1}^+,e_{n}^-), n\in\Z$.
Moreover,  states of $H$ and zeros of $\P^+(0,z)$  belong to the set
$\mZ=\{i\R\}\cup\C_-\cup\bigcup g_n^c\ss \cZ$.

\no iii) A point $z\in\mZ_0$
  is a zero of $\P_0^+$ iff  $z\in\mZ_0\cap \s_{st}(H)$. In particular,

  1) $z\in \ol\C_+\cap \cZ$ is a bound state of $H$,

   2) $z\in \ol\C_-\cap \cZ$  is a resonance of $H$.

\end{lemma}
\no {\bf Proof.} i)
Lemma \ref{Tm} and identity \er{Ptp} yield that each $\P^\pm(x,\cdot )\in \mA(\m_n), x\ge 0$.

Using \er{Tm-2} we deduce that $\P_0^\pm(\m_n)\ne 0$,
then the functions $\P^\pm(x,\cdot), 1/\P_0^\pm\in \mA(\m_n)$. Thus $\m_n$ is not a  state of $H$ and $\m_n$ is a simple zero of $F$.

 ii) The conformal mapping $k(\cdot)$ maps each interval $(e_{n-1}^+,e_{n}^-),
n\ge 1$ onto $(\pi (n-1), \pi n)$. Recall that $\m_n\in [e_{n}^-,e_{n}^+]$.  Then the function $m_\pm$ is analytic on $(e_{n-1}^+,e_{n}^-)$ and $\Im m_\pm(z)\ne 0$
for all $z\in (e_{n-1}^+,e_{n}^-)$. Then  the identity \er{Ptp}
gives $\P^\pm(0,z)\ne $ for all $z\in (e_{n-1}^+,e_{n}^-)$.

Then the  identity $\P^+(0,z)=D(z^2)$ and standard arguments (similar to the case $p=0$, see \cite{K1})
imply that states of $H$ and zeros of $\P^+(0,z)$  belong to the set
$\mZ=\{i\R\}\cup\C_-\cup\bigcup g_n^c\ss \cZ$.

 iii)  The statement  iii) follows from the identities  \er{Ptp},  \er{R}.
$\BBox$

We consider the properties of the states of $H$ which coincide with
unperturbed states $z_n^0$.

\begin{lemma}
\lb{T32}
Let $\z=\m_n+i0\in g_n^+$ or $\z=\m_n-i0\in g_n^-$  for some $n\ge 1$,
where $g_n\ne \es$. Then

\no i) $\wt \vp(0,\m_n)=0$ iff $\F(n_t,\m_n)=0$, where $n_t=\inf_{n\in \N, n\ge t} n$.

\no ii) Let in addition $\z=z_n^0\in \s_{st}(H_0)$. Then $\P_0^-\in \mA(\z)$
and each $\P^+(x,\cdot), x>0$ has a simple pole at $\z$ and there are two cases:

\no 1) if $\wt\vp(0,\m_n)=0$, then $\P_0^+\in\mA(\z), \ \P_0^-(\z)\ne 0$
and $\z\in \s_{st}(H)$. In particular,
\[
\lb{T32-0}
if \qq \z=\m_n+i0\in g_n^+ \qq \Rightarrow \qq \P_0^+(\z)\ne 0,\qq
F(\m_n)=0,\qqq (-1)^nF'(\m_n)>0;
\]
2) if $\wt\vp(0,\m_n)\ne 0$,  then
\begin{multline}
\lb{T32-1}
{\P^+(x,\cdot)\/\P_0^+}\in \mA(\z),x\ge 0, \ , \ \ \P_0^-(\z)\ne 0, \ \
\P_0^+(z)={c_n+O(\ve)\/\ve}\wt\vp(0),\\ \z\notin \s_{st}(H),\qqq
F(\m_n)\ne 0.
\end{multline}

\no iii) $\z\in \s_{bs}(H) $ (or $\z\in \s_{vs}(H) $ ) iff
$\z\in \s_{bs}(H_0) $ (or $z_1\in \s_{vs}(H_0)$ and $\F(n_t,\m_n)=0$.

\no iv) Let $\z\in \s_{st}(H_0)\cap \s_{st}(H)$, then
the same number but on another sheet is not a state of $H$.

\no v) Let $\z\in \s_{st}(H_0)$ and let the same number $\z_1=\ol\z$ but on another sheet is a state of $H$. Then $\z\notin \s_{st}(H)$.

\end{lemma}
\no{\bf Proof.} i) Comparing \er{T31-1} and \er{Ptp} we deduce that
 $\wt \vp(0,\m_n)=0$ iff $\F(0,\m_n)=0$.

ii) Lemma \ref{Tm} yields $m_-\in \mA(\z)$
and each $\P^+(x,\cdot), x>0$ has a simple pole at $\z$
and $m_+(z)={\pm c_n\/\ve}+O(1)$ as $\ve\to 0,\ \ve \in \C_\pm$ and $c_n<0$.

1) If $\wt\vp(0,\m_n)=0$, then \er{Ptp} yields     $\P_0^+\in\mA(\z), \ \P_0^-(\z)\ne 0$.
Thus $\z\in \s_{st}(H)$, since each $\P^+(x,\cdot), x>0$ has a simple pole at $\z$.

Consider the case $\z=\m_n+i0\in g_n^+$. Recall that \er{T31-1} gives
$$
\P^\pm(0,z)=e^{\pm ik(z)n_t}w_\pm(z),\ \ \  w_\pm(z)=\F'(n_t,z)-m_\pm(z)\F(n_t,z),
\qqq n_t=\inf_{n\in \N, n\ge t} n,\  z\in \cZ,
$$
which yields
$$
w_-(\z)=\F'(n_t,\m_n)\ne 0,\qq
w_+(z_1)=\F'(n_t,\m_n)(1-c_n{\pa_z\F(n_t,\m_n)\/\F'(n_t,\m_n)})\ne 0,
$$
since  $\F'(n_t,\m_n)\ne 0, \F(n_t,\m_n)=0$  and $\F'(n_t,\m_n)\pa_z\F(n_t,\m_n)>0$  (see \cite{PT}).
This yields \er{T32-0}, since $(-1)^n\pa_z \vp(1,\m_n)>0$
(see \cite{PT}).

2) Lemma \ref{Tm} yields $m_+(z)={c_n+O(\ve)\/\ve}$ as $\ve=z-\z\to 0$.
Using \er{Ptp}, we obtain
$$
\P_0^+(z)={c_n+O(\ve)\/\ve}\wt\vp(0),\qq
{\P^+(x,z)\/\P_0^+(z)}={\ve \wt\vt(x)-c_n\wt\vp(x)+O(\ve)\/\ve \wt\vt(0)-c_n\wt\vp(0)+O(\ve)}=
{\wt\vp(x)\/\wt\vp(0)}+O(\ve)\qq \as \ \ve\to 0,
$$
since $c_n\wt\vp(0,\m_n)\ne 0$, where $\wt\vp(x)=\wt\vp(x,z),..$. This yields $\vp(1,z)\P_0^+(z)=\pa_z\vp(1,\m_n)c_n+o(1)$ and
$m_-\in \mA(\z)$ gives $\P_0^+(\z)=\wt\vt(0)\ne 0$, which yields $F(\m_n)\ne 0$
and \er{T32-1}.

Using   i) and  ii) we obtain iii).

iv) $\P_0^+\in \mA(\z_1)$
and each $\P^+(x,\cdot)\in \mA(\z_1), x>0$.
Due to ii) $\wt\vp(0,\m_m)=0$, then we obtain $\P_0^+(\z_1)\ne 0$.
Thus $\z_1\notin \s_{st}(H)$.

\no v) Assume that $\z\in \s_{st}(H)$. Then iv) gives contradiction.
Thus $\z\notin \s_{st}(H)$.
\BBox

Consider virtual states, which coincide with the points $e_n^\pm$.

\begin{lemma} \lb{T33}
Let $\z=e_n^-$ or $\z=e_n^+$ for some $n\ge 1$, where $e_n^-<e_n^+$ and let $\ve=z-\z$.

\no i) Let $\z\ne \m_n$ and let $\P_0^+(\z)=0$. Then
 $\z$ is a simple zero of $F$, $\z\in \s_{vs}(H)$
 and
\[
\lb{T33-1}
\P_0^+(z)=\wt\vp(0,\z)c\sqrt \ve+O(\ve),\qq
\mR(x,z)={\P^+(x,z)+O(\ve)\/\wt\vp(0,\z)c\sqrt \ve},\qq c\wt\vp(0,\z)\ne 0.
\]
\no ii) Let $\z=\m_n$ and $\wt\vp(0,\z)\ne 0$. Then
$F(\z)\ne 0$ and each $\mR(x,\cdot), x>0$ has not singularity at $\z$,
and $\z\notin \s_{vs}(H)$.

\no iii) Let $\z=\m_n$ and $\wt\vp(0,\z)=0$. Then $\z\in \s_{vs}(H)$,
$\P_0^\pm(\z)\ne 0$ and $\z$ is a simple zero of $F$ and each $\mR^2(x,\cdot), x>0$ has  a pole at $\z$.
\end{lemma}

\no {\bf Proof.}
i) Lemma \ref{Tm} gives $m_\pm(z)=m_\pm(\z)+c\sqrt \ve+O(\ve)$ as $\ve=z-\z\to 0, c\ne 0$.
We have two cases: 1) Firstly, let $\wt\vp(0,\z)\ne 0$. Then identity \er{Ptp} implies \er{T33-1}.

2) Secondly, if $\wt\vp(0,\z)=0$, then \er{Ptp} implies
 $\P_0^+(\z)=\wt\vt(0,\z)\ne 0$, which gives contradictions.

\no ii) If $\z=\m_n$, then Lemma \ref{Tm} gives $m_\pm(z)=\pm {c\/\sqrt \ve}+O(1), \ve\to 0, c\ne 0$. Then  \er{Ptp} implies
$$
\P_0^\pm(z)=\pm{\wt\vp(0,\z)c\/\sqrt \ve}+O(1),\qq
{\P^+(x,z)\/\P_0^+(z)}={\wt\vt(x,z)+({c\/\sqrt \ve}+O(1)) \wt\vp(x,z)\/{\wt\vp(0,\z)c\/\sqrt \ve}+O(1)}={1+O(\sqrt \ve)\/\wt\vp(0,\z)}.
$$
Thus the function $\mR(x,\cdot), x>0$ has not  singularity at $\z$ and $\z\notin \s_{vs}(H)$, $F(\z)\ne 0$.

iii) If $\wt\vp(0,\z)=0$, then \er{Ptp} gives $\P_0^+(\z)=\wt\vt(0,\z)\ne 0$, since $\wt\vt(0,\z)\ne 0$ and $\b(\z)=0$. Moreover, we obtain
$
\P^+(x,z)=\wt\vt(x,z)+({c\/\sqrt \ve}+O(1)) \wt\vp(x,z),
$
and the function $\mR^2(x,\cdot), x>0$ has a pole at $\z$, $\z\in \s_{vs}(H)$
and $F(\z)=0$.
$\BBox$

\begin{lemma}
\lb{T34}
Let $\l\in \g_n, \l\ne \m_n^2$ be an eigenvalue of $H$ for some $n\ge 0$ and let $z=\sqrt \l\in i\R_+\cup \cup_{n\ge 1} g_n^+$. Then
\[
\lb{T34-1}
C_\l=\int_0^\iy |\P^+(x,z)|^2dx=-{{\P^+}'(0,z)\/2z} \pa_z\P^+(0,z)>0,
\]
\[
\lb{T34-2}
{i2\sin k(z)\/\vp(1,z)}=\P^-(0,z_1){\P^+}'(0,z)\ne 0,\qqq
i\sin k(z)=-(-1)^n\sinh h,\ \ h>0,
\]
\[
\lb{T34-3}
C_\l={(-1)^nF'(z)\sinh h\/ z\vp^2(1,z)\P^-(0,z)^2}>0,
\qqq {(-1)^nF'(z)\/z}>0.
\]
\end{lemma}
\no {\bf Proof.} Using the identity $\{{\pa\/\pa z} \P^+,\P^+\}'=2z(\P^+)^2$
we obtain \er{T34-1}.

Using the Wronskian for the functions $\P^+, \P^-$ and \er{f1} we obtain  $\P^-(0,z){\P^+}'(0,z)=m_+(z)-m_-(z)$,
which yields \er{T34-2}, since $k(z)=\pi n+ih$ for some $h>0$, see the definition
of $k(\cdot)$ before \er{f1}. Then identities \er{T34-1}, \er{T34-2} imply \er{T34-3}.
$\BBox$

\section {Proof of main Theorems}
\setcounter{equation}{0}

{\bf Proof of Theorem \ref{T1}}. i) Asymptotics \er{T1-1}  were proved in Lemma \ref{T21}.

ii) and iii) of Lemma \ref{T32} give \er{st2} for the case
of non-virtual states, i.e.,  $\ne e_n^\pm$.

Lemma \ref{T33} implies \er{st2} for the case
of virtual states.

Lemma \ref{T311} gives \er{st1} for the case
of non-virtual states.
Lemma \ref{T33} implies  \er{st1} for the case
of virtual states.

ii) Using ii) and iii) of Lemma \ref{T32} we obtain \er{ajf}.

iii) Due to i) $\z$ is a zero of $F$, then \er{T22-3} yields \er{T1-2}.
 Lemma \ref{T22} (ii) completes the proof of iii).
\BBox

\no {\bf Proof of Theorem \ref{T2}}.
i) Let $g_n\ne \es$. The entire function $F=\vp(1,\cdot)\P_0^+\P_0^-$
has different sign on $\s_n$ and $\s_{n+1}$, since
$\P_0^+(z)\P_0^-(z)=|\P_0^+(z)|^2>0$ for $z$ inside $\s_n\cup \s_{n+1}$ (see \er{T21-1}) and $\vp(1,\cdot)$ has one simple zero in each interval $[e_n^-,e_n^+]$.
Then $F$ has an odd number of zeros on $[e_n^-,e_n^+]$.

By Lemma \ref{T311}-\ref{T33}, $\z\in g_n^c$ is a state of $H$ iff $\z\in \ol g_n$ is a zero of $F$
(according to the multiplicity). Then the number of states on $g_n^c$ is odd.

  Using Lemma \ref{T22} and \ref{T311} we deduce that there exists an exactly one simple state $z_n$ in each interval $[e_n^-,e_n^+]$  for $g_n\ne \es$ and for  $n\ge 1$ large enough. Moreover, asymptotics $e_n^\pm=\pi n+{p_0+o(1)\/2\pi n}$, see \er{ape} give
\[
\lb{rae}
z_n=\pi n+{p_0+o(1)\/2\pi n}.
\]
Using arguments proving \er{T21-2} we obtain the identities
\[
\lb{T21-20}
\wt \vt(0,z)=1+\int_0^t\vp(x,z)q(x)\wt \vt(x,z)dx,\qqq     \wt \vp(0,z)=\int_0^t\vp(x,z)q(x)\wt \vp(x,z)dx,
\]
The standard iteration procedure and \er{rae} give the asymptotics
\[
\lb{asvt1}
\wt \vt(0,z_n)=1+O(1/n),\qqq     ,
\]
\[
\lb{asvp1}
\wt\vp(0,z_n)=\int_0^t{\sin^2 z_nx\/z_n^2}q(x)dx+O(1/n^3)={q_0-\wh q_{cn}+O({1\/n})\/2(\pi n)^2},
\]
where $\wh q_{cn}=\int_0^tq(x)\cos 2\pi nxdx$.
Using \er{LD0} and $\P^\pm=\wt\vt+m_\pm\wt\vp$, see \er{Ptp},
 we obtain
\[
\lb{idF1}
F=F_1+F_2+F_3,\qqq F_1=\vp(1,\cdot)\wt\vt_0^2,\qq F_2=2\b\wt\vt_0\wt\vp_0,\qqq F_3=-\vt'(1,\cdot)\wt\vp_0^2,
\]
where for shortness $\wt\vt_0=\wt\vt(0,z),\ \wt\vp_0=\wt\vp(0,z)$.
Using estimates \er{asvt1}, \er{asvp1}, we obtain
\begin{multline}
\lb{T31-4}
F_1(z_n)=\vp(1,z_n)(1+O(n^{-1})),\qqq \qqq F_3(z_n)=O(n^{-4})
\\
F_2(z_n)=(-1)^n{(p_{sn}+O({1\/n}))\/\pi n} {(q_0-\wh q_{cn}+O({1\/n}))\/2(\pi n)^2}=f_n+O(n^{-4})\qq \as \ n\to \iy,
\end{multline}
where $f_n=(-1)^n{p_{sn}(q_0-\wh q_{cn})\/2(\pi n)^3}$. Combine these asymptotics and the identity $F(z_n)=0$ we get
\[
\vp(1,z_n)=-F_2(z_n)+O(n^{-4})=-f_n+O(n^{-4}).
\]
Then, using $\vp(1,z_n)=\pa_z\vp(1,\m_n)\d_n+O(n^{-4})$,
where $z_n=\m_n+\d_n$, we obtain
$$
\pa_z\vp(1,\m_n)\d_n=-f_n+O(n^{-4})
$$
and the asymptotics $\pa_z\vp(1,\m_n)={(-1)^n+O({1\/n})\/(\pi n)}$ give
$$
\d_n=-{f_n\/\pa_z\vp(1,\m_n)}+O(n^{-3})=-{(\wh q_0-\wh q_{cn})p_{sn}+O({1\/n})\/2(\pi n)^2},
$$
which yields \er{T2-1}.

 Denote by $\cN^+(r,f)$ the number of zeros of $f$ with real part $\geq 0$ having modulus $\leq r$,
and by $\cN^-(r,f)$ the number of its zeros with real part $< 0$
having modulus $\leq r$, each zero being counted according to its multiplicity.
We recall the well known result (see [Koo]).

\no    {\bf Theorem (Levinson).}
{\it  Let the entire function $f\in \mC^\r$.
Then $  \cN^{\pm }(r,f)={r\/ \pi }(\r+o(1))$ as $r\to \iy$,
and for each $\d >0$ the number of zeros of $f$ with modulus $\leq r$
lying outside both of the two sectors $|\arg z | , |\arg z -\pi |<\d$
is $o(r)$ for $r\to \iy$.}

Let $\cN (r,f)$ be the total number of  zeros of $f$  with modulus $\le r$.
Denote by $\cN_+(r,f)$ (or $\cN_-(r,f)$) the number of zeros of $f$ with imaginary part $>0$ (or $<0$)
having modulus $\le r$, each zero being counted according to its multiplicity.

Let $s_0=0$ and $\pm s_n>0, n\in \N$  be all real zeros of $F$
and let $n_0$ be the multiplicity of the zero $s_0=0$.
Define the entire function $F_1=z^{n_0}\lim_{r\to \iy}\prod_{0<s_n\le r}(1-{z^2\/ s_n^2})$. The Levinson Theorem and Lemma \ref{T21} imply
\[
\cN(r,F)=\cN(r,F_1)+\cN(r,F/F_1)=2r{1+2t+o(1)\/\pi},\qq
\cN(r,F_1)=2r{1+o(1)\/\pi},
\]
\[
\cN_-(r,F)=\cN_+(r,F)=\cN_-(r,\P_0^+)-N_0,
\]
as $r\to\iy$, where $N_0$ is the number of non-positive eigenvalues of $H$. Thus
\[
2\cN_-(r,F)=2r{2t+o(1)\/\pi},
\]
which yields \er{T2-2}.

ii) Using Lemma \ref{T34} we obtain the statements ii) and iii).
\BBox

{\bf Proof of Theorem \ref{T3}.}
Let $z=e_n^\pm$. Identity \er{T31-1} and $k(e_n^\pm)=\pi n$ yield
\[
\lb{Jep}
\P_0^-(z)=\P_0^+(z)=(-1)^Nw_+(z),\ \ \  w_+(z)=\F'(n_t,z)-{\b(z)\/\vp(1,z)}\F(n_t,z),\qq
N=n_tn.
\]
 Estimates \er{efs1} and $e_n^\pm=\pi n+\ve_n(p_0\pm |p_n|+O(\ve_n)), \ \ve_n={1\/2\pi n}$ (see \er{ape}) give
\[
\F'(n_t,z)=(-1)^N+{O(1)\/n},\qq  \F(n_t,z)={\sin n_tz\/\pi n}+{O(1)\/n^2}={O(1)\/n^2}.
\]
Using \er{efs}, we obtain
$$
\sin e_n^\pm=(-1)^n\sin { \pm |p_n|+O({1\/n})\/2\pi n}=(-1)^n{ \pm|p_n|+O({1\/n})\/2\pi n},
$$$$
\vp(1,e_n^\pm)={\sin e_n^\pm\/\pi n}+{(-1)^np_{cn}\/2\pi^2n^2} +{O(1)\/n^3}=
(-1)^n{\pm|p_n|+p_{cn}+O({1\/n})\/2\pi^2n^2}.
$$
Then the estimate $\sqrt{x^2+y^2}-y\ge x$ for $y,x\ge 0$ gives
$|p_n|\pm p_{cn}\ge |p_{sn}|$, which yields
\[
\lb{bdvp}
{\b(e_n^+)\/\vp(1,e_n^+)}=\pi n{p_{sn}+O({1\/n})\/\pm|p_n|+p_{cn}+O({1\/n})}=
O(\pi n),
\]
since $|p_{sn}|\ge {1\/n^\a}$. Combining \er{Jep}-\er{bdvp} and \er{abn}, we obtain
$\P_0^+(e_n^\pm)=1+o(1)$. The function $\P_0^+(z)$ is analytic on $g_n^-$
and $\P_0^+(e_n^\pm)=1+o(1)$. Thus $\P_0^+(z)$ has not
 zeros on $g_n^-$, since by Theorem \ref{T2}, the function
 $F$ has exactly 1 zero on each $\ol g_n\ne \es$ at large $n>1$.

Let $\m_n+i0\in g_n^+$ be a bound state of $H_0$ for some  $n$ large enough.
Then Lemma \ref{Tm} implies $h_{sn}>0$. Moreover,
\er{anc} gives $h_{sn}=-{p_{sn}+O({1\/n})\/2\pi n}$ as $n\to \iy$.
Thus $p_{sn}<-{1\/n^\a}$ at large $n>1$ and asymptotics \er{T2-1} gives
that the bound state  $z_n>\m_n$ if $q_0>0$ and $z_n<\m_n$ if $q_0<0$.
The proof of other cases is similar.
\BBox

{\bf Proof of Theorem \ref{T4}.} i)
 Using the identities \er{T21-2} and \er{Ptp} we obtain
\[
\lb{iii}
\P_0^+=Y_1+{i\sin k\/\vp_1}\wt \vp(z_n),\qq Y_1=\wt \vt(z_n)+{\b\/\vp_1}\wt \vp(z_n).
\]
Note that \er{LD0} gives $\b(\m_n)=0$. Then asymptotics \er{T2-1}, \er{asb},
\er{efs1} imply
\[
{\b(z_n)\/\vp(1,z_n)}={\b'(\m_n)+O({\ve_n^3})\/\pa_z\vp(1,\m_n)+O({\ve_n^3})}
=o(1)\qqq \as \qq n\to \iy,\qqq \ve_n={1\/2\pi n}
\]
where we used asymptotics $\pa_z\vp(1,\m_n)={(-1)^n+O({\ve_n})\/(\pi n)}$
and $\b'(\m_n)={o(1)\/n}$. Thus \er{asvt1}, \er{asvp1} give
\[
Y_1=1+O(\ve_n), \qqq \wt\vp(0,z_n)=2\ve_n^2(b_n+O(\ve_n)), \qqq b_n=q_0-\wh q_{cn} 
\]

Below we need the identities and the asymptotics as $n\to \iy$ from \cite{KK}:
\[
\lb{35} (-1)^{n+1}i\sin k(z)=\sinh v(z)=\pm |\D^2(z)-1|^{1\/2}>0\qq
\ all \qq z\in  g_n^\pm,
\]
\[
\lb{pav}
v(z)=\pm |(z-e_n^-)(e_n^+-z)|^{1\/2}(1+O(n^{-2})),\qq
\sinh v(z)=v(z)(1+O(|g_n|^2),\qq  z\in \ol g_n^\pm.
\]
We rewrite the equation $\P_0^+=0$ in the form $\vp_1Y_1=-i\sin k\wt \vp(z_n)$
Then we obtain
$$
2\d \ve_n(1+O(\ve_n))=v(z)2\ve_n^2(b_n+O(\ve_n))=\sqrt{\d(|g_n|-\d)}2\ve_n^2(b_n+O(\ve_n)),
$$
$$
\sqrt \d =\ve_n\sqrt{|g_n|-\d}(b_n+O(\ve_n)), \qqq \d=z_n-\m_n,
$$
where $\sqrt \d>0$ if $b_n>0$  and $\sqrt \d<0$ if $b_n<0$. Then last
asymptotics imply $\d =\ve_n^2|g_n|(b_n+O(\ve_n))^2$, where $b_n=q_0-\wh q_{cn}, \ve_n={1\/2\pi n}$, which yields \er{T4-1}.

ii) Let $q\in \cQ_t, q_0=0$ and let each  $|\wh q_{cn}|>n^{-\a}$
for some $\a\in (0,1)$ and for $n$ large enough.
The proof of other cases is similar.
Then using the inverse spectral theory from \cite{K5}, see page 3, for any sequence  $\vk=(\vk_n)_1^\iy\in \ell^2, \vk_n\ge 0$
there exists a potential $p\in L^2(0,1)$  such that each gap length  $|\g_n|=\vk_n, n\ge 1$ for $n$ large enough.
Moreover, for $n$ large enough we can the gap in the form $\g_n=(E_n^-,E_n^+)$, where $\m_n^2=E_n^-$ or $\m_n^2=E_n^+$.
In order to choose $\m_n^2=E_n^-$ or $\m_n^2=E_n^+$ we do the following.
For any sequence $\s=(\s_n)_{1}^\iy$, where $\s_n\in \{0,1\}$,
using Theorem \ref{T3} (i) we obtain:

If $\s_n=1$ and $\wh q_{cn}<-n^{-\a}$ (or $\wh q_{cn}>n^{-\a}$),   then taking $\m_n^2=E_n^-$ (or $\m_n^2=E_n^+$) we deduce that $\l_n$ is an eigenvalue
for $n$ large enough.

If $\s_n=0$ and $\wh q_{cn}>n^{-\a}$ (or $\wh q_{cn}<-n^{-\a}$), then taking $\m_n^2=E_n^-$ (or $\m_n^2=E_n^+$) we deduce that $\l_n$ is an
antibound state for $n$ large enough.

iii) Let $p\in L^2(0,1)$ such that $\g_n=(E_n^-,E_n^+)$, where $\m_n^2=E_n^-$ or $\m_n^2=E_n^+$  for $n\in \N_0$ large enough.
Let $\s=(\s_n)_{1}^\iy$ be any sequence, where $\s_n\in \{0,1\}$.
We take $\wh |q_{cn}|>n^{-\a}$ for $n\in \N_0$ large enough. We need to choose
the sign of $q_{cn}$. Using Theorem \ref{T3} (i) we take the sign of $q_{cn}$
by

If $\s_n=0$ and $\l_n^0=E_n^-$ (or $\l_n^0=E_n^+$), then taking
$\wh q_{cn}>n^{-\a}$ (or $\wh q_{cn}<-n^{-\a}$) we deduce that
$\l_n$ is an antibound state.

If $\s_n=1$ and $\l_n^0=E_n^-$ (or $\l_n^0=E_n^-$), then taking
$\wh q_{cn}<-n^{-\a}$ (or $\wh q_{cn}>n^{-\a}$) we deduce that
$\l_n$ is an eigenvalue.
\BBox

\no {\bf Proof of Corollary  \ref{T5}.}
Let $\#_{bs}(H,\O)$ (or $\#_{abs}(H,\O)$)  be the total number of bound states
 (or anti bound states) of $H$ on the segment $\O\ss \g_n^{(1)}$ (or $\O\ss \g_n^{(2)}$) for some $n\ge 0$. Here each state being counted according to its multiplicity.

Recall that $H_\t=H_0+q_\t$, where $q_\t=q({x\/\t})$ and $\t\to \iy$.
Let $\O=[E_1,E_2]\ss \ol \g_n^{(1)}$ for some $n\ge 0$.
Then using the result of Sobolev \cite{So} with a modification
of Schmidt \cite{Sc} we obtain
\[
\lb{asbs}
\#_{bs}(H_\t,\O)={\t}\int_0^\iy \rt(\r(E_2-q(x))-\r(E_2-q(x))\rt)dx+o(\t)\qq \as \qq \t\to \iy.
\]
Theorem \ref{T2} (iii) implies $\#_{abs}(H_\t,\O^{(2)})\ge 1+\#_{bs}(H_\t,\O)$,
which together with \er{asbs} yield \er{T5-1}.
\BBox

\no {\bf Acknowledgments.}
\setlength{\itemsep}{-\parskip} \footnotesize
The research was partially supported  by EPSRC grant EP/D054621.
The various parts of this paper were written at  ESI, Vienna, Universit\'e de Gen$\grave{\rm e}$ve, Section de Mathematiques and
Mathematical Institute of the Tsukuba Univ., Japan.
The author is grateful to the Institutes for the hospitality.
The author would like also to thank A. Sobolev (London) and K. Schmidt (Cardiff) for useful discussions about the asymptotics associated with Corollary \ref{T5}.


\end{document}